\documentstyle[preprint,aps,epsf]{revtex}
\newcommand{\beq}{\begin{equation}}
\newcommand{\eeq}{\end{equation}}
\newcommand{\ds}{\displaystyle}
\newcommand{\beqar}{\begin{eqnarray}}
\newcommand{\eeqar}{\end{eqnarray}}

\begin{document}
\tightenlines
\draft

\title{
 Local equilibrium in heavy ion collisions. \\
 Microscopic model versus statistical model analysis.
}
\author{
L.~V.~Bravina,$^{1,2}$ E.~E.~Zabrodin,$^{1,2,}$\footnote{Present 
address: Institute for Theoretical Physics, University of 
T\"ubingen, D-72076 T\"ubingen, Germany.} 
M.~I.~Gorenstein,$^{1,3}$ S.~A.~Bass,$^{4}$
M.~Belkacem,$^{1}$ M.~Bleicher,$^{1}$ M.~Brandstetter,$^{1}$ 
C.~Ernst,$^{1}$ M.~Hofmann,$^{1}$ L.~Neise,$^{1}$ 
S.~Soff,$^{1,5}$ H.~Weber,$^{1}$
H.~St{\"o}cker,$^{1}$ and W.~Greiner$^{1}$
}
\address{
$^1$Institute for Theoretical Physics, University of Frankfurt,
Robert-Mayer-Strasse 8-10, D-60054 Frankfurt, Germany \\ 
$^2$Institute for Nuclear Physics, Moscow State University,
119899 Moscow, Russia \\
$^3$Bogolyubov Institute for Theoretical Physics, Kiev, Ukraine \\
$^4$Department of Physics, Duke University, Durham, 
North Caroline 27708-0305, USA \\
$^5$Gesellschaft f\"ur Schwerionenforschung, Darmstadt, Germany \\
}

\maketitle

\begin{abstract}
The assumption of local equilibrium in relativistic heavy ion 
collisions at energies from 10.7{\it A} GeV (AGS) up to 160{\it A} GeV 
(SPS) is checked in the microscopic transport model. Dynamical 
calculations performed for a central cell in the reaction are compared 
to the predictions of the thermal statistical model. 
We find that kinetic, thermal, and chemical equilibration of the 
expanding hadronic matter are nearly approached late in central
collisions at AGS energy for $t \geq 10$ fm/$c$ in a central cell.
At these times the equation of state may be approximated by a 
simple dependence $P \cong (0.12-0.15)\, \varepsilon$.
Increasing deviations of the yields and the energy spectra of hadrons 
from statistical model values are observed for increasing energy,
40{\it A} GeV and 160{\it A} GeV. These violations of local 
equilibrium indicate that a fully equilibrated state is not reached, 
not even in the central cell of heavy ion collisions at energies 
above 10{\it A} GeV. The origin of these findings is traced to the 
multiparticle decays of strings and many-body decays of resonances.
\end{abstract}


\widetext

\section{Introduction} 
\label{sec1}

Since the beginning of the 1930s, when cosmic ray cascades of various 
particles were detected, physicists are trying to describe the process 
of multiple production of particles in ultrarelativistic collisions 
of hadrons and nuclei. The idea of Fermi, namely that all secondary 
particles produced in a Lorentz-contracted volume are in statistical 
equilibrium \cite{Fer50}, was further modified by Pomeranchuk 
\cite{Pom51}. It was finally developed by Landau into the 
hydrodynamic theory \cite{Land53,LaBe56} of multiparticle processes.
The most important quantitative predictions of the hydrodynamic 
theory, such as the dependence of the average particle multiplicity 
$\langle N \rangle $ on total energy of the system $\sqrt{s}$, the 
rapidity distributions $d N / d y$, violation of Feynman scaling, the 
mean value of the transverse momentum $\langle p_\perp \rangle$ and 
its dependence on $\sqrt{s}$ and the mass of the secondaries, have 
been verified experimentally. On the other hand, the basic assumptions 
of the theory, such as the formation of the initial state, the 
relaxation rate of the system to local equilibrium (LE), the sharpness 
of the freeze-out (FO) and, finally, the equation of state (EOS) of 
hot and dense hadronic matter, are based on rough estimates, which 
have not been rigorously proven yet.

To answer these questions, one can analyze the dynamics provided by 
microscopic Monte Carlo simulations, i.e., microscopic string, 
cascade, transport, etc. models. These models
describe experimental data on hadronic and nuclear collisions in a
wide energy range reasonably well, but do not postulate local 
equilibrium. Consequently, the EOS and macroscopic variables like 
temperature, entropy or chemical potential are not implied, but can
be calculated, if the system does actually reach LE.
For instance, the hypothesis of sharp freeze-out of secondaries in 
heavy ion collisions was checked by means of three microscopic Monte 
Carlo models: QGSM \cite{Bra95}, RQMD \cite{Sor96} and UrQMD 
\cite{UrQMD}.
It was found that these models do neither exhibit a thin or a thick
freeze-out layer, which resemble the hyperbolic surface predicted 
by Bjorken scaling model \cite{Bjor83}.

The present paper employs the recently developed
ultrarelativistic quantum molecular dynamics (UrQMD) model 
\cite{UrQMD,Blei98} to examine the approach to local equilibrium of 
hot and dense nuclear matter, produced in central heavy ion collisions 
at energies from AGS to SPS. Local (and sometimes, in fireball models,
even global) equilibrium is the basic {\it ad hoc\/} assumption of the 
macroscopic hydrodynamic models. It is
usually assumed that the nonequilibrium initial stage of nuclear
collisions, during which shock waves, partonic jets, etc., heat the 
system, is considerably shorter than the characteristic hadronization 
times. Evidently, there must be dissipative and irreversible processes 
leading to equilibration. One may adopt the scheme of binary 
collisions, in which the correlation of the many-particle distribution
functions, describing highly nonequilibrium states, rapidly sets in
\cite{Bogo46}. The typical time scale for these processes are 
collision times, $\tau_{\rm kin} \approx \tau_{\rm coll}$. Then a 
kinetic stage is approached, where the $N$-body distribution functions 
are reduced to many one-particle distribution functions, one for each
particle species. For time scales sufficiently larger than 
$\tau_{\rm coll}$, the evolution of the system can be described in 
terms of local average particle number, their velocities and energies. 
These local average values are the moments of the one-particle 
distribution functions, and the hydrodynamic stage arises. Other 
processes, which can cause even faster equilibration, are collective 
instabilities, convective turbulent transport or chaotic relaxation 
\cite{StGr86,Carr84}. They can yield a 
crude estimate of the relaxation times in the system. On the other 
hand, multiparticle processes can lead to a delay in reaching 
equilibrium, because the produced particles are not thermalized.
The time scale of the equilibration may appear too long as compared 
to the typical hadronization times.
 
\begin{figure}[htp]
\centerline{\epsfysize=17cm \epsfbox{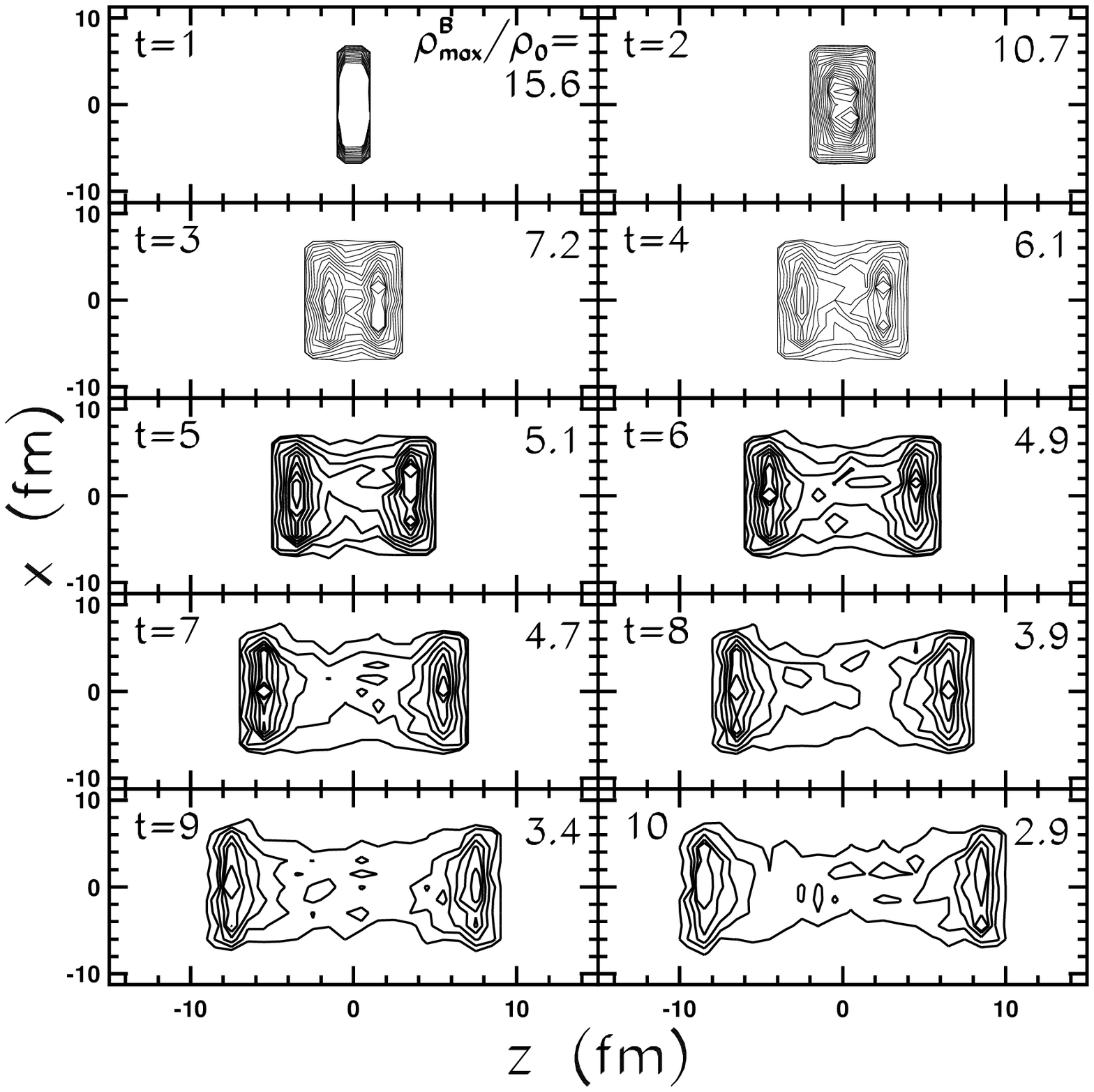}}
\caption{
Space-time evolution picture $d^2N/dxdz$ for baryonic densities
in central $-0.75 < \Delta y < 0.75$~fm slice obtained in Pb+Pb 
collisions at 160{\it A} GeV with zero impact parameter.
Contours show the baryonic densities $\rho_B=0.5\rho_0$, $1\rho_0$, 
etc.
}
\label{fig1}
\end{figure}

As emphasized in \cite{LV98plb}, according to an UrQMD analysis,
it is unlikely that {\it global} thermal equilibrium sets in for 
central Au+Au collisions at the AGS energy. This statement remains 
true at higher energies also.  Figure~\ref{fig1} depicts the time 
evolution of the baryon density in a single central Pb+Pb collision 
at 160A GeV. Two disks of baryon rich matter, remnants of the 
colliding nuclei, consisting mostly of resonances and diquarks, are 
seen in the fragmentation regions. The volume between the disks 
becomes more and more baryon dilute, but never purely homogeneous. 
Apparently, global equilibrium is not reached even in central Pb+Pb 
collisions at SPS energies. However, the occurrence of LE in the 
central zone of the heavy ion reaction is still not ruled out. 

To verify how close the hot hadronic matter in the cell is to
equilibrated matter, one can do a comparison with the statistical 
model (SM) of a hadron gas \cite{LV98plb,LV98jpg}. Three parameters, 
namely, the energy density $\varepsilon$, the baryon density 
$\rho_{\rm B}$, and the strangeness density $\rho_{\rm S}$, extracted 
from the analysis of the cell, are inserted into the equations for an 
equilibrated ideal gas of hadrons. Then all characteristics of the 
system in equilibrium, including the yields of different hadronic 
species, their temperature $T$, and chemical potentials, 
$\mu_{\rm B}$ and $\mu_{\rm S}$, may be calculated unambiguously. 
If the yields and the energy spectra of the hadrons in the cell
are sufficiently close to those of the SM, one can take this as 
indication for the creation of equilibrated hadronic matter in the 
central reaction zone. 

This method is applied to extract the equation of state (EOS), 
which connects the pressure $P$ and the energy density 
$\varepsilon$ of the system (generally, either $P,\ \varepsilon ,
\ \rho_{\rm B}$ or $P,\ T\ {\rm and}\ \mu_{\rm B}$). 
Note that, without the EOS, the system of relativistic hydrodynamic
equations, which describe the evolution of hadronic matter, is 
incomplete. Therefore, the EOS, as extracted from the microscopic 
model, has a direct impact on the parametrizations used in the 
macroscopic (hydrodynamic) models.

This paper is organized as follows: a brief description of the UrQMD 
model is given in Sec.~\ref{sec2}. Here the necessary and sufficient 
criteria of local equilibrium are formulated also. It is shown that 
the hadron distributions in the UrQMD central cell become isotropic at 
$t \cong 8\,$fm/$c$, irrespective of the initial energy of the 
reaction. This means that {\it kinetic\/} equilibrium is reached.
Section~\ref{sec3} presents the basic equations of the statistical 
model of the ideal hadronic gas, which is applied for the analysis 
of the hadron distributions in the central cell. The relaxation of 
the hadronic matter in the central zone of central heavy ion 
collisions to thermal and chemical equilibrium is studied in 
Sec.~\ref{sec4}. The UrQMD cell and the SM results are compared for 
center-of-mass energies of 10.7{\it A} GeV (Au+Au, AGS), 40{\it A} 
GeV (Pb+Pb, SPS) and 160{\it A} GeV (Pb+Pb, SPS).
Finally, the conclusions are drawn in Sec.~\ref{sec5}.

\section{
Criteria of local equilibrium and conditions in the UrQMD cell}
\label{sec2}

\subsection{The UrQMD model}
\label{subs2a}

The UrQMD is a microscopic transport model designed for the
description of hadron-hadron ($hh$), hadron-nucleus ($hA$)
and nucleus-nucleus ($AA$) collisions for energies spanning a few 
hundred MeV up to hundreds of GeV per nucleon in the
center-of-mass system (c.m. system). The model is presented in  
detail in \cite{UrQMD,Blei98}. As discrete, quantized degrees of 
freedom, the model contains 55 baryon and 32 meson states, together 
with their antiparticles and explicit isospin-projected states, with 
masses up to 2.25 GeV/$c^2$. The tables of the experimentally 
available hadron cross sections, resonance widths and decay modes are 
implemented as well.

At moderate energies the dynamics of $hh$ or $AA$ collisions is
described in UrQMD in terms of interactions between the hadrons and 
their excited states (resonances). At higher values of the 
four-momentum transfer, $hh$ interactions cannot be reduced to the 
hadron-resonance picture anymore, and new excited objects, color 
strings, come into play. The excitation of the strings make it 
energetically favorable to break the string into pieces by producing 
multiple $q \bar q$-pairs from the vacuum. Due to the fact that the 
color string is assumed to be uniformly stretched, the hadrons 
produced as a result of the string fragmentation will be uniformly 
distributed in rapidity between the endpoints of the string.
The propagation of the hadrons is governed by Hamilton equations of 
motion, with a binary collision term of the form of relativistic 
Boltzmann-Uehling-Uhlenbeck (BUU) transport model.
The Pauli principle is taken into account via the blocking of the 
final state, if the outgoing phase space is occupied.
No Bose enhancement effects are implemented in the model yet.

\begin{figure}[htp]
\centerline{\epsfysize=15cm \epsfbox{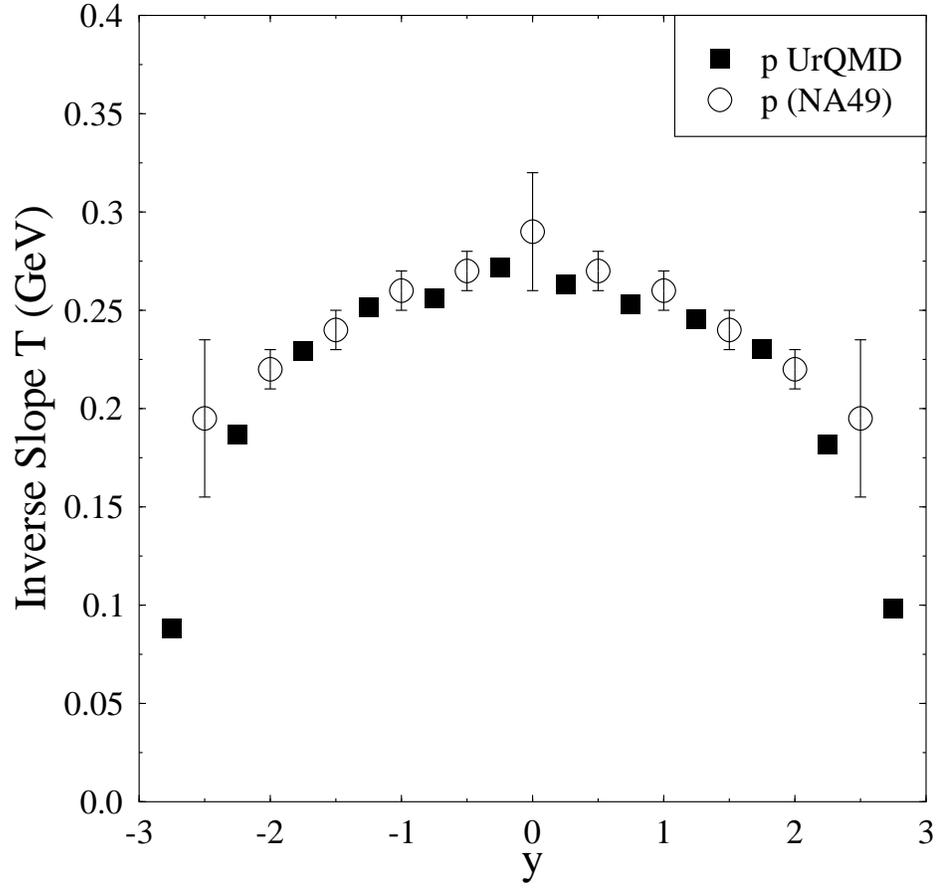}}
\caption{
Apparent ``temperature," $T$, of protons as a function of 
rapidity, $y$, extracted from the analysis of central ($b \leq 3$
fm) Pb+Pb collisions at 160{\it A} GeV (circles). Squares indicate 
the UrQMD calculations.
}
\label{fig2}
\end{figure}

In its present version \cite{URL1.1} UrQMD describes the main 
properties of both hadronic and nuclear interactions 
\cite{UrQMD,Blei98} reasonably well. Very important for the further 
analysis is the fact that the UrQMD model reproduces the experimental 
transverse mass spectra of hadrons in different rapidity intervals, 
as shown in Fig.~\ref{fig2}. The inverse slope parameter, extracted 
from the Boltzmann fit to the spectra, is directly linked to the 
temperature in the statistical model. Thus, if the equilibrium 
conditions (see below) are satisfied, one may estimate the 
temperature of the cell. 

\subsection{Preequilibrium stage}
\label{subs2b}

Let us first define our system. The sizes of the central zone of
perfectly central (impact parameter $b = 0\,$fm) heavy ion collisions
should be neither too small nor too large. The effects caused by the 
collective flow of particles also can wash out the equilibration
picture. In order to diminish number of distorting factors we choose
a central cubic cell of volume $V = 5\times 5\times 5 = 125\,$fm$^3$
centered around the origin of the c.m. system of colliding nuclei.
Due to the absence of the preferable direction of the collective
motion, the collective velocity of the cell is essentially zero.

Then, to decide whether or not the local equilibrium in the cell is 
reached, one has to introduce criteria of the equilibrium. In 
statistical physics the equilibrium state is defined as a state with
maximum entropy \cite{LaLi80}. However, the direct calculation of the
entropy evolution in the cell is notoriously difficult. The cell is
not an isolated system. Particles leave it freely, and neither 
internal energy nor particle number is conserved. Therefore we should
apply other, more suitable for this problem, criteria of equilibrium
bearing in mind that they are consequences of the general principle 
of maximum entropy. 

We start from the necessary conditions which characterize the 
preequilibrium stage in the central cell. The flow velocities there
should have minimum gradient tending to zero. It means that the
local equilibrium in the cell cannot be reached earlier then certain
time $t^{\rm cross} = 2 R /(\gamma_{\rm c.m.} v_{\rm c.m.})$ during 
which the Lorentz contracted nuclei would have pass through each 
other. Here $R$ is the radius of the nuclei, and the times 
$t^{\rm cross}$, typical for each reaction, are 5.46 fm/$c$ 
(10.7{\it A} GeV), 2.9 fm/$c$ (40{\it A} GeV), and 1.44 fm/$c$ 
(160{\it A} GeV). After $t = t^{\rm cross}$ the collective flow of 
freely streaming hadrons rapidly vanishes.

It looks likely that for the cell of small longitudinal size the 
pre-equilibrium stage sets in very quickly. For instance, for the 
central $4\times 4\times 1$ fm$^3$ cell in Pb+Pb collisions at SPS one 
may expect the equilibration already at $t^{\rm eq} = t^{\rm cross} + 
\Delta z /2 \beta \cong 2$ fm/$c$. But the detailed analysis shows 
\cite{LV98jpg} that the hadronic matter in the different central 
cells equilibrates at the same rate which does not depend on the 
longitudinal size of the cell. 

\begin{figure}[htp]
\centerline{\epsfysize=17cm \epsfbox{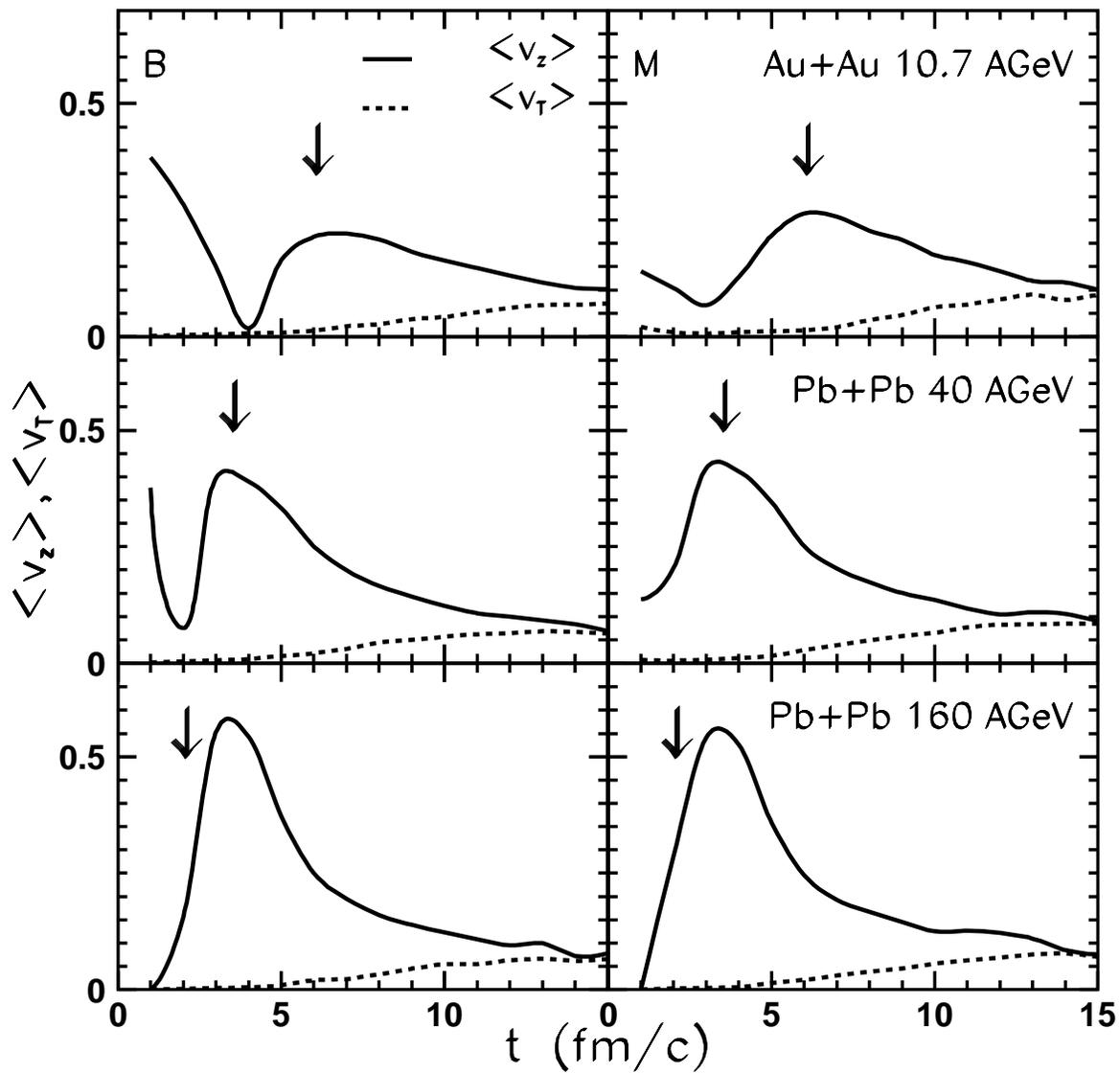}}
\caption{
Average
longitudinal (solid line) and transverse (dashed line) velocities of
baryons (left panels) and mesons (right panels) as functions of time
in asymmetric cell $0 \leq \{x,y,z\} \leq 2.5$ fm for three
reactions in question. Vertical arrows denote the crossing time
$t^{\rm cross}$ (see text). 
}
\label{fig3}
\end{figure}

Figure~\ref{fig3} presents the average transverse and longitudinal
flow of hadrons in 1/8-th of the central cell with the coordinates
$0 \leq \{x,y,z\} \leq 2.5$ as a function of time $t$.
The longitudinal flow reaches its maximum value at times from 
$t = 4$ fm/$c$ (SPS) to $t = 6$ fm/$c$ (AGS). Then it drops and
converges to the transverse flow. At $t = 10$ fm/$c$ the 
longitudinal component of the collective flow in the central cell is
about 0.1--0.15$\,c$ only. This corresponds to the temperature
distortion $T_{\rm fl}\approx m_N\cdot v_{\rm fl}^2/3 \leq 7$ MeV.
Disappearance of the flow implies
{\bf (i)} isotropy of the velocity distributions, which leads to
{\bf (ii)} isotropy of the diagonal elements of the pressure tensor, 
calculated from the virial theorem \cite{Bere92},
\beq
\ds
P_{\{x,y,z\}} = {1 \over {3V}}\, \sum_{i=h}
\frac{p^2_{i\{x,y,z\}}}{(m_i^2~+~p_i^2)^{1/2}} ,
\label{eq1}
\eeq
containing the volume of the cell $V$ and the mass and the momentum
of the $i$th hadron, $m_i$ and $p_i$, correspondingly.

\begin{figure}[htp]
\centerline{\epsfysize=17cm \epsfbox{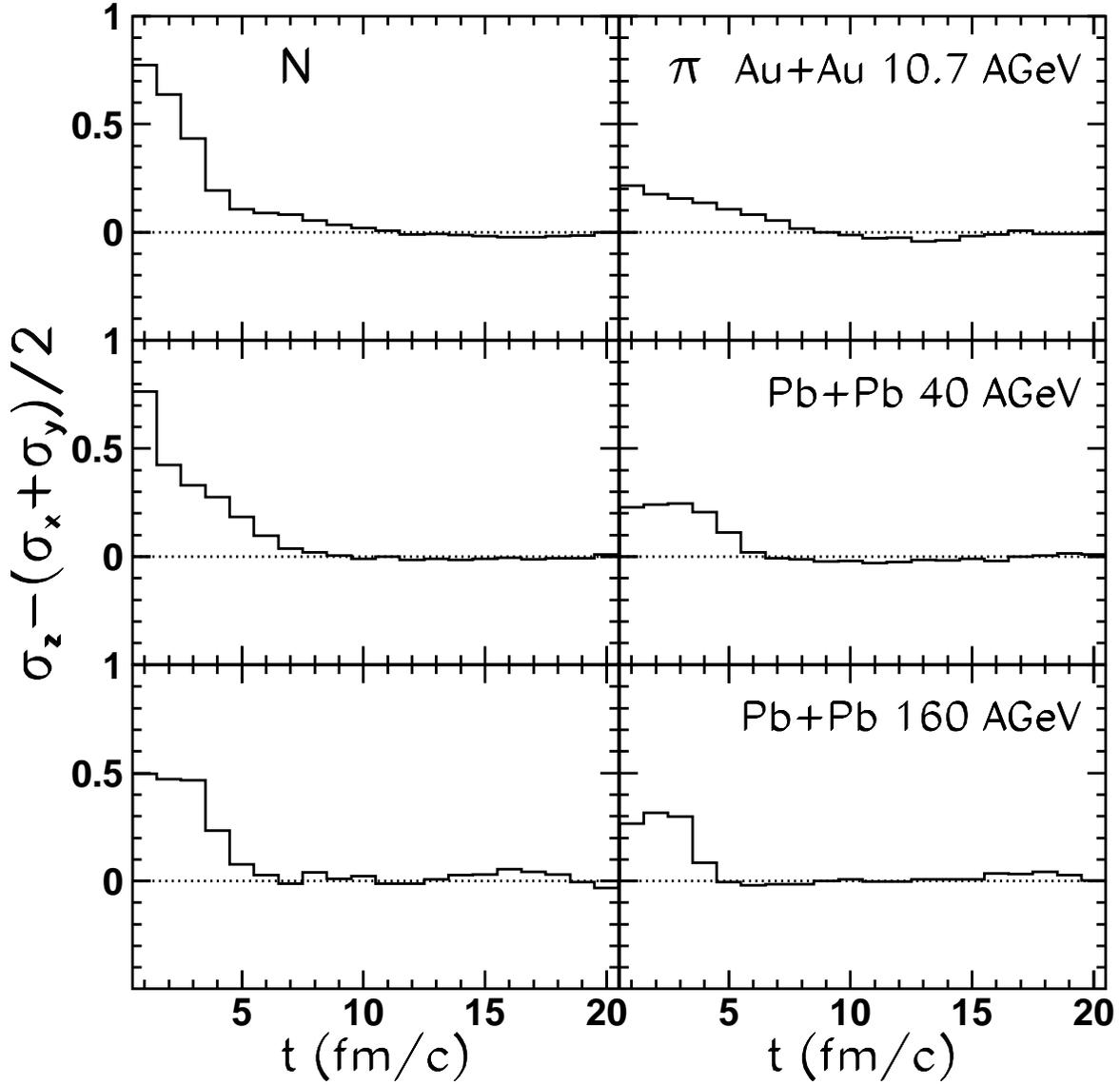}}
\caption{
Anisotropy function 
$f_a^{(2)} = \sigma_z^{(2)} - (\sigma_x^{(2)} + \sigma_y^{(2)}) / 2$
of the velocity distributions of nucleons (left panels) and pions 
(right panels) for 10.7{\it A} GeV (upper row), 40{\it A} GeV 
(middle row), and 160{\it A} GeV (lower row), respectively.
}
\label{fig4}
\end{figure}

\begin{figure}[htp]
\centerline{\epsfysize=17cm \epsfbox{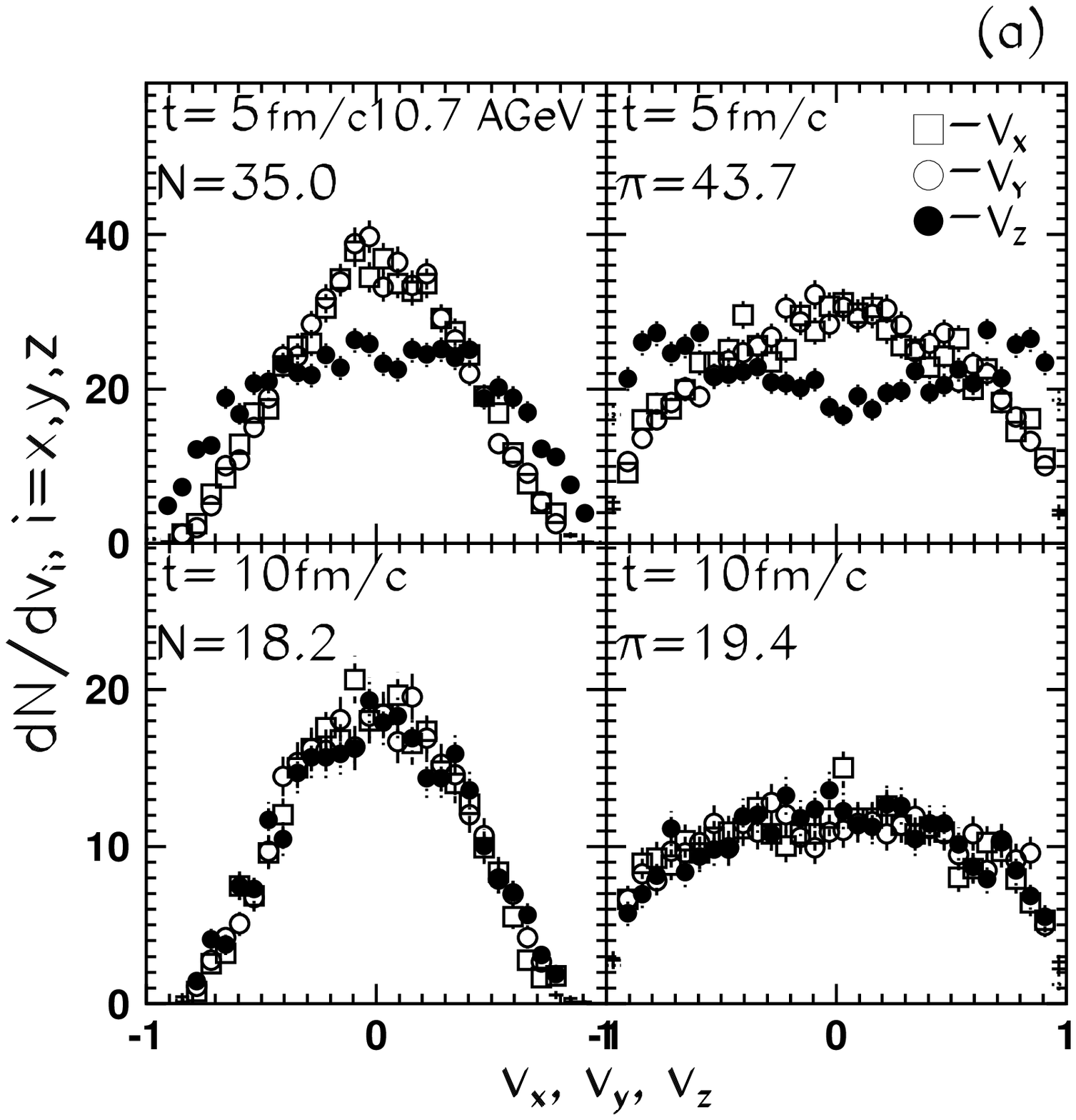}}
\caption{
(a) Nucleon (left frame) and pion (right frame) velocity distributions 
$dN/dv_i$ [$i=z$ ($\bullet $), $x$ ($\Box$) and $y$ ($\bigcirc$)] in 
central cell of volume 125 fm$^3$ in Au+Au collisions at 10.7{\it A} 
GeV at $t=$5 fm/$c$ (upper frame) and at 10 fm/$c$ (lower frame). \\
(b) The same as (a) but for Pb+Pb collisions at 40{\it A} GeV
at $t=$6 fm/$c$ (upper frame) and at 8 fm/$c$ (lower frame). \\ 
(c) The same as (a) but for Pb+Pb collisions at 160{\it A} GeV
at $t=$5 fm/$c$ (upper frame) and at 8 fm/$c$ (lower frame). 
}
\centerline{\epsfysize=17cm \epsfbox{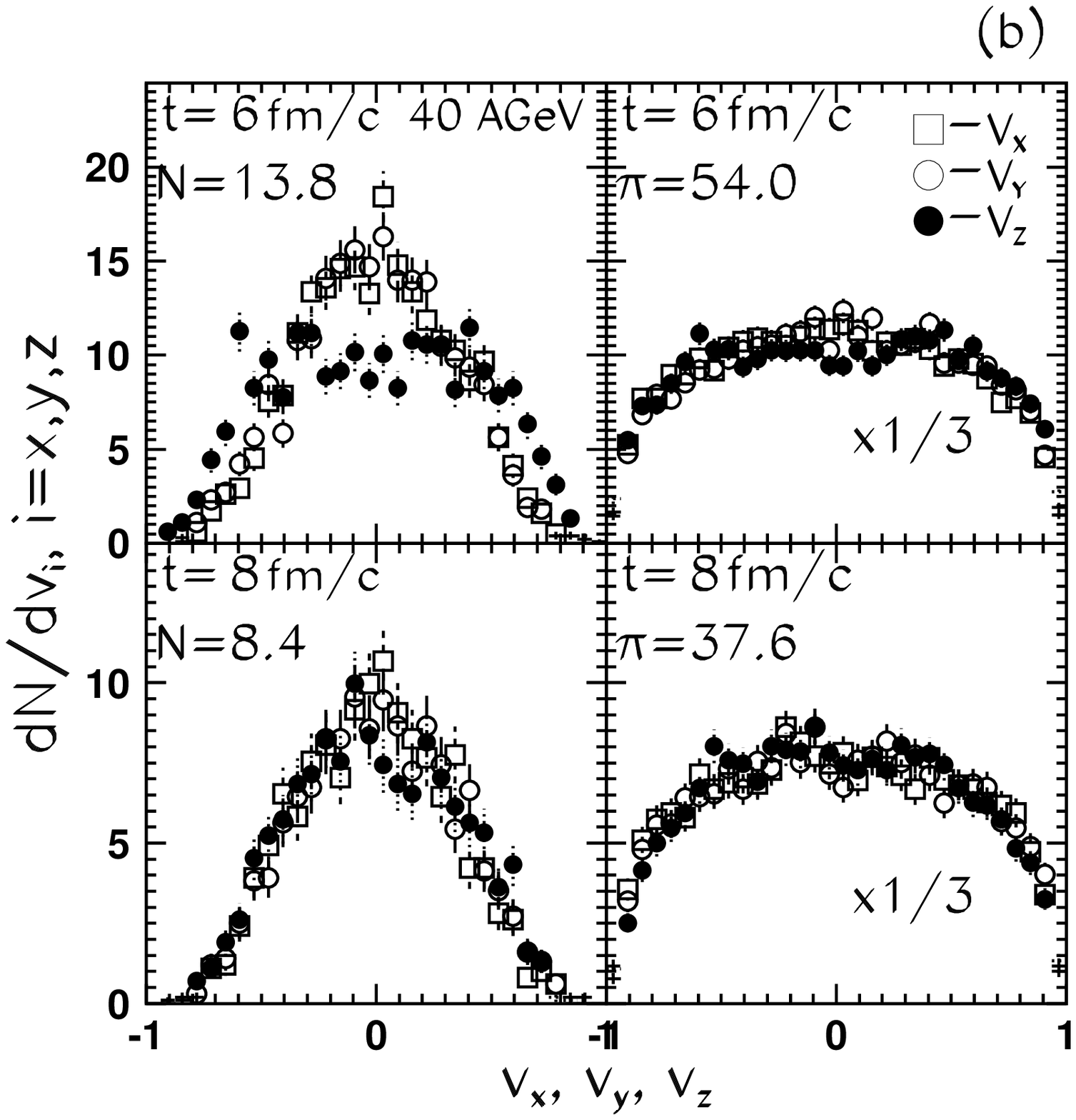}}
\centerline{\epsfysize=17cm \epsfbox{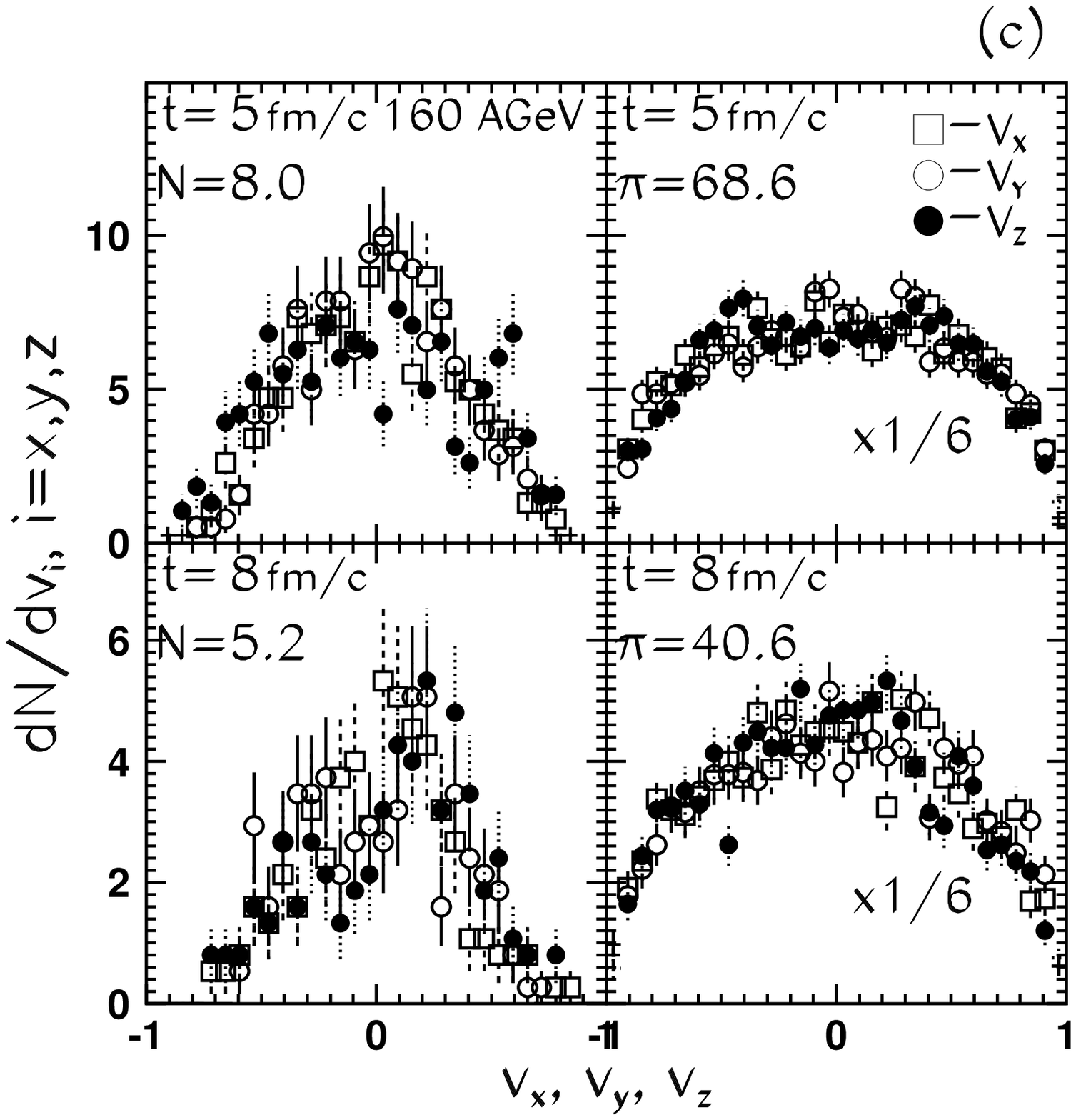}}
\label{fig5}
\end{figure}

The method of moments of the distribution function is a useful tool
to investigate irregularities in the particle spectra in high energy
physics. Indeed, it is possible to make a conclusion about isotropy 
of the velocity distributions in terms of the function 
\beq
\ds
f_a^{(n)} = \sigma_z^{(n)} - \frac{1}{2} \left( \sigma_x^{(n)} +
\sigma_y^{(n)} \right) ,
\label{eq2}
\eeq
where $\sigma_{x,y,z}^{(n)}$ is the $n$th moment of the $x$, $y$,
or $z$ distribution. The function $f_a^{(n)}$ is a measure of 
anisotropy of the particle average velocities in longitudinal and
transverse directions. The results of the calculations of particle
velocity anisotropy of nucleons and pions produced in the central cell 
at 10.7{\it A} GeV, 40{\it A} GeV and 160{\it A} GeV are given in 
Fig.~\ref{fig4} for the function $f_a^{(2)}$. It seems that the 
velocity distributions 
becomes isotropic already at $t = 5 - 7\,$fm/$c$. But in equilibrium 
at energies and temperatures in question hadrons should have 
Maxwellian, or normal, velocity distribution, which is the only one
satisfying the principle of maximum entropy. The $dN/dv$
distributions of nucleons and pions for all three energies
are shown in Fig.~\ref{fig5}. One can see that despite the
isotropy of collective velocities, the shapes of the distributions 
become close to the normal one at $t = 8 - 10 $fm/$c$, not earlier. 
Therefore, the second moments of the velocity distribution functions 
are insufficient to study the system anisotropy, and one should apply
the moments of higher order. 

\begin{figure}[htp]
\centerline{\epsfysize=17cm \epsfbox{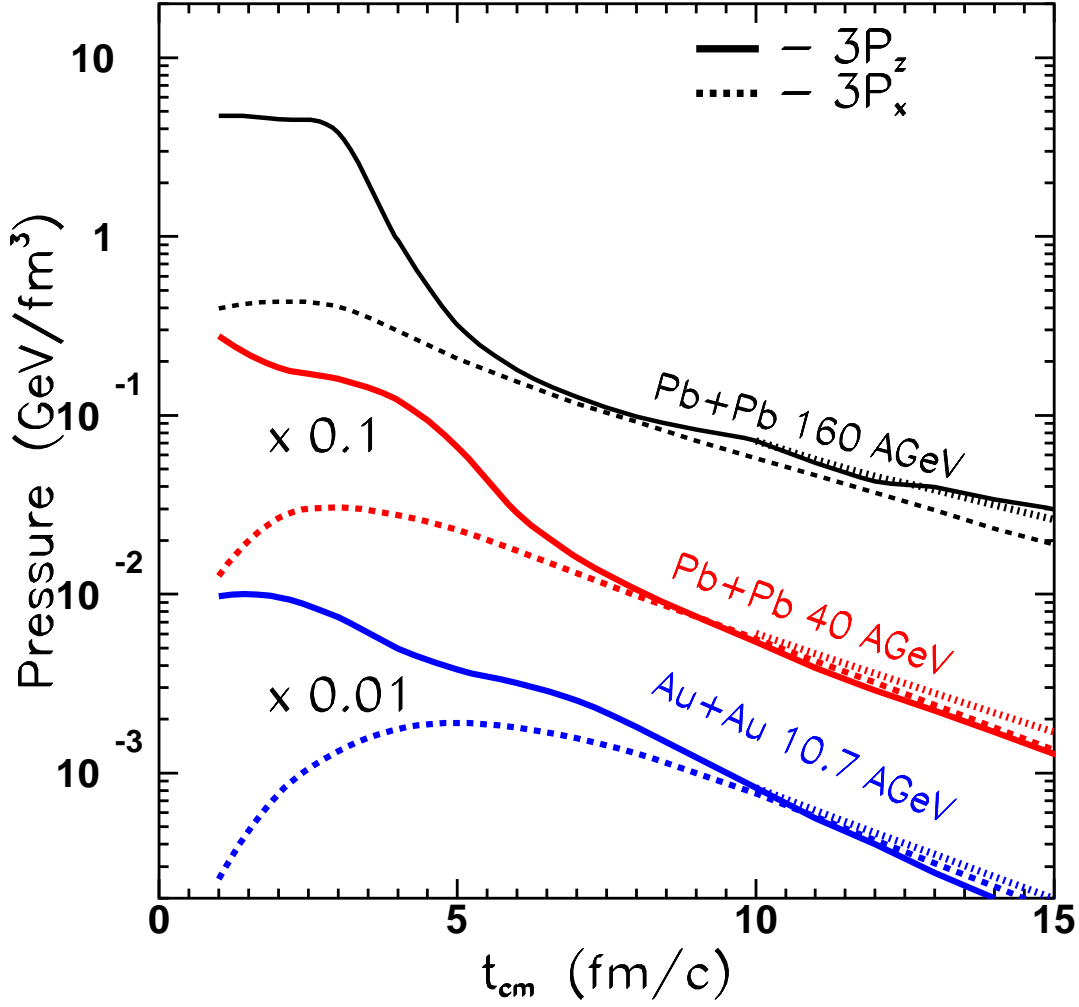}}
\caption{
The longitudinal ($3 P_{\{z\}}$, solid curves) and the 
transverse ($3 P_{\{x\}}$, dashed curves) diagonal components 
of the microscopic pressure tensor in the central 125 fm$^3$ cell 
of Au+Au and Pb+Pb collisions at 10.7, 40, and 160{\it A} GeV 
calculated from the virial theorem Eq.~(\protect\ref{eq1}). Dotted 
curves indicate the pressure given by the statistical model
Eq.~(\protect\ref{eq10}).
}
\label{fig6}
\end{figure}

The requirement of isotropy of the velocity distributions is 
closely related to the requirement of the pressure isotropy.
Here the momentum distributions of particles are integrated over the
whole number of hadron species, as given by Eq.~(\ref{eq1}).
Figure~\ref{fig6} depicts the time evolution of the pressure in
longitudinal and transverse directions calculated for all three
energies in question. These components become very close (within the
5\%-limit of accuracy) to each other also at $t \geq 8$ fm/$c$. The 
result does
not depend practically on the initial energy of colliding nuclei.

It is worth noting that the appearance of the preequilibrium
stage does not imply inevitably that the matter in the cell would
be in equilibrium forever. The preequilibrium stage in the cell holds
for a period of about $8-10$ fm/$c$ (Fig.~\ref{fig6}). Then the 
system develops again the anisotropy in the pressure and velocity
sectors due to significant reduction of number of collisions between 
particles. 

\subsection{Criteria of thermal and chemical equilibrium}
\label{subs2c}

Suppose conditions {\bf (i)} and {\bf (ii)} are fulfilled.
Does it mean that the hadronic matter in the cell is in a state
close to the equilibrium? No, because both criteria concern
the {\it kinetic\/} preequilibrium stage rather than the 
{\it thermal\/} equilibrium one. {\it Kinetic\/} equilibrium implies
isotropy of the velocity distributions of particles (and, therefore,
isotropy of pressure) together with the requirement that their 
velocity distributions must be Maxwellian. {\it Thermal\/} 
equilibrium indicates that the macroscopic characteristics of the
system are nearly equal to their average values. 
In thermal equilibrium the system is characterized by a unique 
temperature $T$. Then, the principle of maximum entropy compels the 
particles of mass $m_i$ to obey the equilibrium distribution function
\beq
\ds
F(p,m_i) = \left[ \exp{(\sqrt{p^2 + m_i^2} -\mu)/T} \pm 1 
 \right]^{-1} .
\label{eq3}
\eeq
Here $p$ is the momentum of particle, $\mu$ is its chemical potential,
$``+"$ sign stands for fermions and $``-"$ for bosons. In the case 
$\exp{[(E_i - \mu)/T]} \gg 1$, where $E_i^2 = p^2 + m_i^2$, the 
Maxwell velocity distribution follows automatically from 
Eq.~(\ref{eq3}). Therefore, 
if the number of particles is conserved, then both definitions of 
{\it kinetic\/} and {\it thermal\/} equilibrium are equivalent 
\cite{LaLi80,deGr80}. But in strong interactions at high energies
(as well as in chemical reactions) the number of interacting particles 
and their origin is changed.
Thus, the condition of {\it thermal\/} equilibrium should be
completed by the requirement of {\it chemical\/} equilibrium. Both
criteria read {\bf (iii)} the distribution functions of hadrons are
close to the equilibrium distribution functions, given by 
Eq.~(\ref{eq3}) (thermal equilibration), and {\bf (iv)} the yields 
of hadrons become saturated (chemical equilibrium) after a certain 
period. The latter condition assumes that any inverse reaction 
proceeds with the same rate as the direct reaction. This means,
particularly, that the chemical potentials of nonconserved charges
vanish, and the chemical potential $\mu_j$, assigned to a given 
particle $j$, is simply 
\beq
\ds
\mu_j = \mu_{\rm B} B_j + \mu_{\rm S} S_j ,
\label{eq3a}
\eeq
where $B_j$ and $S_j$ denote the baryon charge and strangeness of the 
particle, and $\mu_{\rm B},\ \mu_{\rm S}$ is the baryo- and
strangeness chemical potential, respectively.

After the preequilibrium conditions are satisfied, one may address 
the question on {\it thermal\/} and {\it chemical\/} equilibrium in 
the cell. But how can we apply the thermostatic criteria
to the dynamic picture in the cell, where the internal parameters are
instantly changing? The standard procedure \cite{LV98plb,LV98jpg}
is to compare the snapshot of particle yields and spectra in the cell 
at given time with those predicted by the statistical thermal model of 
hadron gas \cite{SOG81,GMQY94,CEST97}. If these spectra are close to
each other the hadronic matter in the cell is assumed to reach thermal 
and chemical equilibrium. The simplicity of the SM has led to a very 
abundant literature (see, e.g. \cite{SBM,BrSt95,Bec96,Belk98} and 
references therein). Therefore, we shall recall briefly some of its 
principle features.

\section{Statistical model of ideal hadron gas}
\label{sec3}

For the further analysis the thermodynamical parameters of the system, 
$T$, $\mu_{\rm B}$ and $\mu_{\rm S}$, at each step of the time 
evolution of the colliding system were extracted from
the predictions of the statistical model of an ideal hadron gas with 
the same 55 baryon and 32 meson species and their antistates
considered in UrQMD model. As an
input the SM uses the total energy density $\varepsilon$, baryon
density $\rho_{\rm B}$ and strangeness density $\rho_{\rm S}$, 
determined within the UrQMD model during the dynamical evolution of 
the central zone with volume $V$ of the A+A system:
\beqar
\ds
\label{eq4}
\varepsilon^{\rm mic} &=& \frac{1}{V}\sum_i
            E_i^{\rm SM}(T,\mu_{\rm B},\mu_{\rm S}), \\
\label{eq5}
\rho_{\rm B}^{\rm mic}&=& \frac{1}{V}\sum_i
            B_i\cdot N_i^{\rm SM}(T,\mu_{\rm B},\mu_{\rm S}), \\
\label{eq6}
\rho_{\rm S}^{\rm mic}&=& \frac{1}{V}\sum_i
            S_i\cdot N_i^{\rm SM}(T,\mu_{\rm B},\mu_{\rm S}).
\eeqar
Here  $B_i$, $S_i$
are the baryon charge and strangeness of the hadron species $i$, 
whose particle yields,  $N_i^{\rm SM}$, and total energy, 
$E_i^{\rm SM}$, are calculated within the SM as:
\beqar
\ds
N_i^{\rm SM} &=& \frac{V g_i}{2\pi^2\hbar^3}\int_0^{\infty}p^2
 f(p,m_i) d p , \\
\label{eq7}
E_i^{\rm SM} &=& \frac{V g_i}{2\pi^2\hbar^3}\int_0^{\infty}
 p^2 \, \sqrt{p^2+m_i^2}\, f(p,m_i) d p ,
\label{eq8}
\eeqar
where $p$, $m_i$ and $g_i$ are the momentum, mass and the degeneracy 
factor of the hadron species $i$. The distribution function 
$f(p,m_i)$ is given by Eq.~(\ref{eq3}) with the chemical potential
$\mu = \mu_{\rm B}B_i + \mu_{\rm S}S_i$. Then,
instead of Fermi-Dirac or Bose-Einstein distributions we use for all
hadronic species the classical Boltzmann distribution function
\beq
\ds
f^{\rm Boltz}(p,m_i)
 = \exp{ \left( - \frac{\sqrt{p^2+m_i^2}-\mu_{\rm B} B_i-\mu_{\rm S}
 S_i}{T} \right)}.
\label{eq9}
\eeq
At temperatures above 100 MeV the only visible difference
(about 10\%) between quantum and classical descriptions is in the
yields of pions \cite{LV98jpg}.

The hadron pressure given by the statistical model reads
\beq
\ds
P^{\rm SM}=~\sum_i \frac{g_i}{2\pi ^2\hbar^3}\int_0^{\infty}
p^2 \frac{p^2}{3(p^2+m_i^2)^{1/2}}~f(p,m_i) d p .
\label{eq10}
\eeq
Finally, the entropy density $s^{\rm SM}$ can be calculated for the 
ideal gas model either by the Gibbs thermodynamical identity
\beq
\ds
\varepsilon^{\rm mic}=T^{\rm SM}s^{\rm SM}+\mu_{\rm B}^{\rm SM}
\rho_{\rm B}^{\rm mic}+\mu_{\rm S}^{\rm SM}\rho_{\rm S}^{\rm mic}-
P^{\rm SM},
\label{eq11}
\eeq
or as a sum over all particles of the product $f(p,m_i)\,[1 - 
\ln{f(p,m_i)}]$ integrated over all possible momentum states
\beq
\ds
s^{\rm SM} = -\sum_i \frac{g_i}{2\pi^2\hbar^3} \int_0^{\infty}
f(p,m_i)\, \left[ \ln{f(p,m_i)}-1 \right] \, p^2 d p.
\label{eq12}
\eeq
According to the principle of maximum entropy the value of 
$s^{\rm SM}$ calculated from Eq.~(\ref{eq11}) or from Eq.~(\ref{eq12})
represents the maximum value of the entropy density in the system for 
a given particle composition and given set of microscopic parameters, 
i.e., energy density, $\varepsilon^{\rm mic}$, baryon density, $\rho_
{\rm B}^{\rm mic}$, and strangeness density, $\rho_{\rm S}^{\rm mic}$.

\section{UrQMD versus statistical model. Results and discussion}
\label{sec4}

\subsection{Baryon density and strangeness in the cell}
\label{subs4a}

As shown in Sec.~\ref{sec2}, the {\it kinetic\/} equilibrium is
attained in the central cell at $t \approx 10$ fm/$c$ for all three
reactions. The fraction of non-formed particles at this time is less
than 20\% and rapidly vanishes. Therefore, $t = 10$ fm/$c$ is chosen
as a starting point of the direct comparison between UrQMD and SM.
Substitution of the values $\{\varepsilon^{\rm mic}, \rho_{\rm B}^{
\rm mic}, \rho_{\rm S}^{\rm mic} \}$, calculated in the cell for all 
three reactions in question at $10 \leq t \leq 18$ fm/$c$, into 
Eqs.~(\ref{eq4})--(\ref{eq6}) gives us the key parameters $T, 
\mu_{\rm B}$ and $\mu_{\rm S}$ needed to reproduce particle spectra 
in the statistical model. The input and output parameters are listed 
in Tables~\ref{tab1}--\ref{tab3}. Apparently, the conditions in the 
cell are different for all reactions even at this late stage of the 
expansion. Because of the different expansion rates the higher
temperatures and lower values of the baryonic chemical potential are
assigned to the system of heavy ions colliding at SPS energy, which
is the highest one in our case. The final time of the calculations 
may be estimated from the imposed usual hydrodynamic freeze-out 
conditions, e.g. $\rho_{tot} \approx 0.5\, \rho_0$ or 
$\varepsilon = 0.1$ GeV/fm$^3$, i.e. 18--20 fm/$c$ for all three 
reactions. At these times the fraction of already frozen particles in 
the central cell is about 40--47 \%, irrespective on the initial
energy of colliding nuclei. 

The baryon density in the central zone of the collision at the late
stage is not larger than 0.15 fm$^{-3}$ for all three reactions.
Note also that
at AGS energies we deal with baryon rich matter, where about 70\% of
the total energy is carried by baryons, while at SPS most of the
energy is deposited in the mesonic sector (more then 70\%). At
40{\it A} GeV the mesonic and baryonic parts of energy are equal.

\begin{figure}[htp]  
\centerline{\epsfysize=15cm \epsfbox{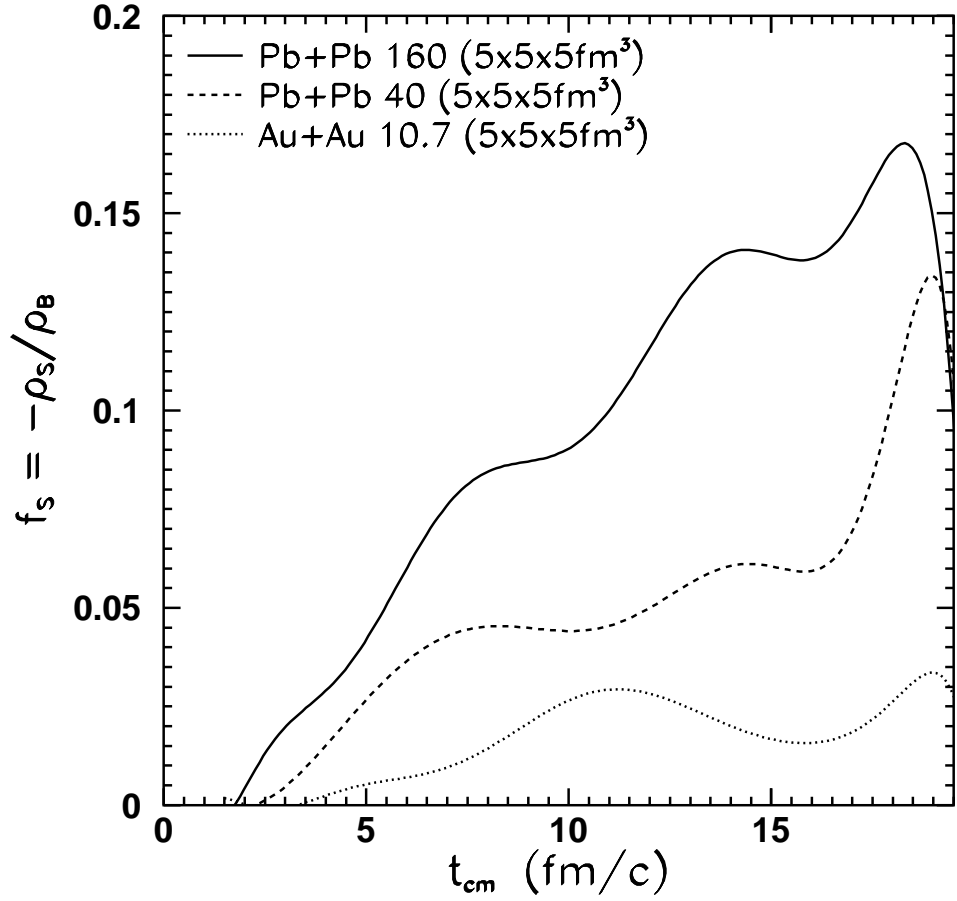}}
\caption{
Time evolution of strangeness per baryon, $f_{s}=-\rho_{\rm S} /
\rho_{\rm B}$, obtained in central cell of volume $V=125$ fm$^3$ in 
heavy ion collisions at 10.7, 40, and 160{\it A} GeV.
}
\label{fig7}
\end{figure}

At all energies from 10.7{\it A} GeV to 160{\it A} GeV the total strangeness 
density of all particles in the central cell
is small and negative. This result is independent on the size of
the cell \cite{LV98jpg}. The origin of this effect is quite simple.
Strange particles are produced in pairs, for instance, kaons are 
produced mainly together with lambdas. The total strangeness of the
reaction is essentially zero but, owing to small interaction cross
section with hadrons, $K$'s are leaving the central cell much earlier 
than $\Lambda$'s or $\overline{K}$'s. The strangeness density has a 
minimum somewhere at the maximum overlap of the nuclei and then it 
relaxes to zero. This evolution behavior of the strangeness density
$\ds \rho_{\rm S}=\sum_{i} S_i\cdot n_i$, where $S_i$ and
$n_i$ are the strangeness and density of the hadron species $i$,
can not be explained by simple combination of e.g. $K$'s, 
$\overline{K}$'s and $\Lambda$'s, but is defined by contributions of 
all species, carrying the strange charge.
For the ratio $f_{s}=-\rho_{\rm S}/\rho_{\rm B}$ (Fig.~\ref{fig7}) the 
behavior is opposite. This ratio rises continuously with time because 
the baryon density decreases much faster than the strangeness density.

At the early stages of the reaction the strange charge is carried
mostly by resonances. At AGS energy the positive strangeness of 
mesons, mostly $K$'s, is compensated by the contribution of both 
baryons (like $\Lambda$'s) and mesons (like $\overline{K}$'s), 
carrying the negative strange charge.
It makes the net strangeness density negative, even though small.
At SPS energy the contribution of strange baryons to the total
strangeness is relatively small, so the difference in strangeness
is defined mostly by meson contributions.

To decide whether the strangeness density in the cell is small or 
not we have also performed calculations for SM with zero strangeness 
density.  As shown in Table~\ref{tab4}, although the hadronic yields 
themselves are only slightly affected by the ``symmetrization" of 
strangeness, their ratios are changed more distinctly. For instance, 
ratio $F_K=K/\overline K$ in the central cell drops from 6.48 to 
5.74 at AGS energy, from 2.96 to 2.60 at 40{\it A} GeV, and from 1.82 to 
1.58 at SPS energy. The total effect is about 15\% for all three 
reactions.
 
\subsection{Isentropic expansion and EOS}
\label{subs4b}

\begin{figure}[htp]  %
\centerline{\epsfysize=17cm \epsfbox{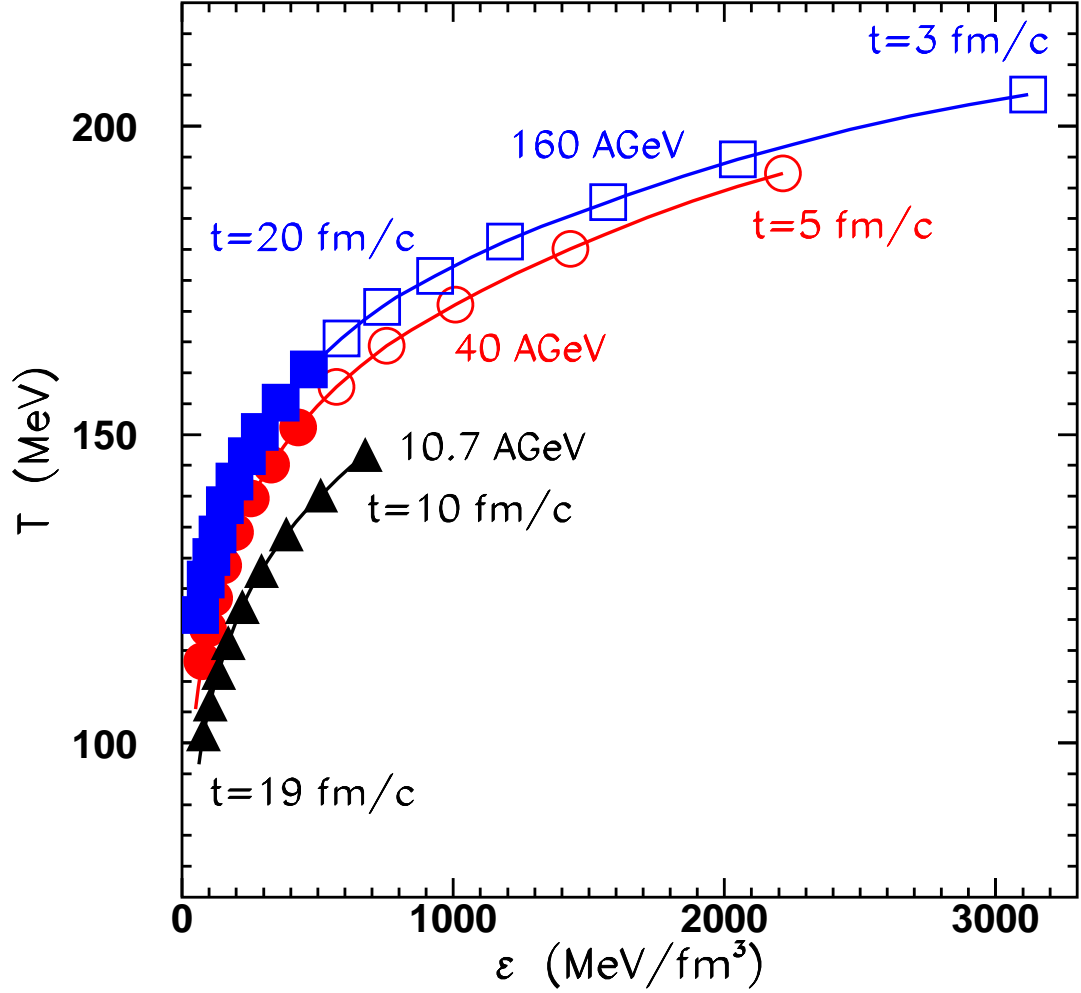}}
\caption{
The evolution of the energy density, $\varepsilon$, and temperature,
$T$, in the central cell of heavy ion collisions at 10.7{\it A} GeV 
(triangles), 40{\it A} GeV (circles), and 160{\it A} GeV (boxes).
$\varepsilon$ is obtained in the UrQMD cell calculations;
$T$ is extracted from the fit to the SM.
}
\label{fig8}
\end{figure}

\begin{figure}[htp]  %
\centerline{\epsfysize=17cm \epsfbox{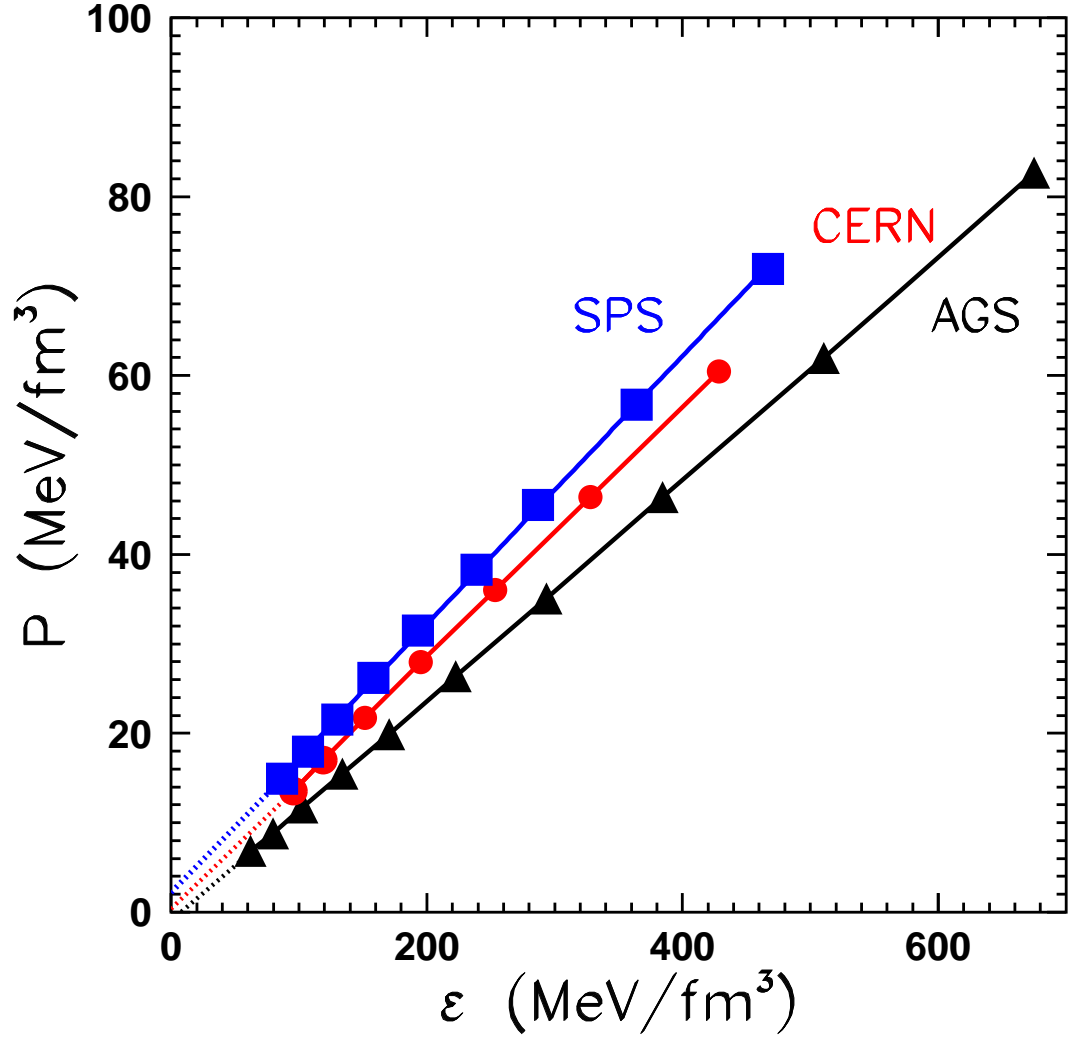}}
\caption{
The same as Fig.\protect~\ref{fig8}, but for the pressure-energy
density ($P$-$\varepsilon$ plane). Pressure is taken from the 
UrQMD cell calculations.
}
\label{fig9}
\end{figure}

Two other important facts may be gained from the 
Tables~\ref{tab1}--\ref{tab3}. The entropy per
baryon ratio in the cell is almost constant and varies from 
$S/A = s/\rho_{\rm B} \cong 12$ at AGS to $s/\rho_{\rm B} \cong 
38 \pm 2$ at SPS energy. This result supports the Landau idea of 
isentropic expansion of a relativistic fluid. The isentropic-like
expansion is demonstrated in Fig.~\ref{fig8} which presents the
evolution of the central cell in the $T-\varepsilon$ plane. 
Also, the fact that the microscopic pressure calculated according to 
Eq.~(\ref{eq1}) is nicely reproduced by the Eq.~(\ref{eq10}) for the
pressure of ideal hadron gas \cite{LV98plb} favors the applicability 
of the hydrodynamic description of relativistic heavy ion collisions. 
The ratio $P/\varepsilon$, shown in Fig.~\ref{fig9}, is constant 
for the whole time interval for all three energies. Thus the 
equation of state, which connects pressure with energy density, has a 
rather simple form $P(\varepsilon) / \varepsilon = 0.12$ (AGS), 0.13 
(40{\it A} GeV) and 0.15 (SPS), where 
$\ds \left( d P / d \varepsilon \right)^{1/2} = c_s$ 
corresponds to the speed of sound in the medium. For the ideal
ultra-relativistic gas $c_s^2 = 1/3$, while the presence of resonances
diminishes the sonic velocity to $c_s^2 \cong 0.14$ \cite{Shur72},
which is in quantitative agreement with our calculations.
It is important to stress that the EOS, extracted from the UrQMD
analysis of the evolution of hot hadronic matter in the central cell
of heavy ion collisions at energies from AGS to SPS, does not contain
any kind of softening, which may be associated with the phase 
transition \cite{HuSh95} between the quark-gluon plasma (QGP) and 
hadronic phase.

\begin{figure}[htp]  %
\centerline{\epsfysize=17cm \epsfbox{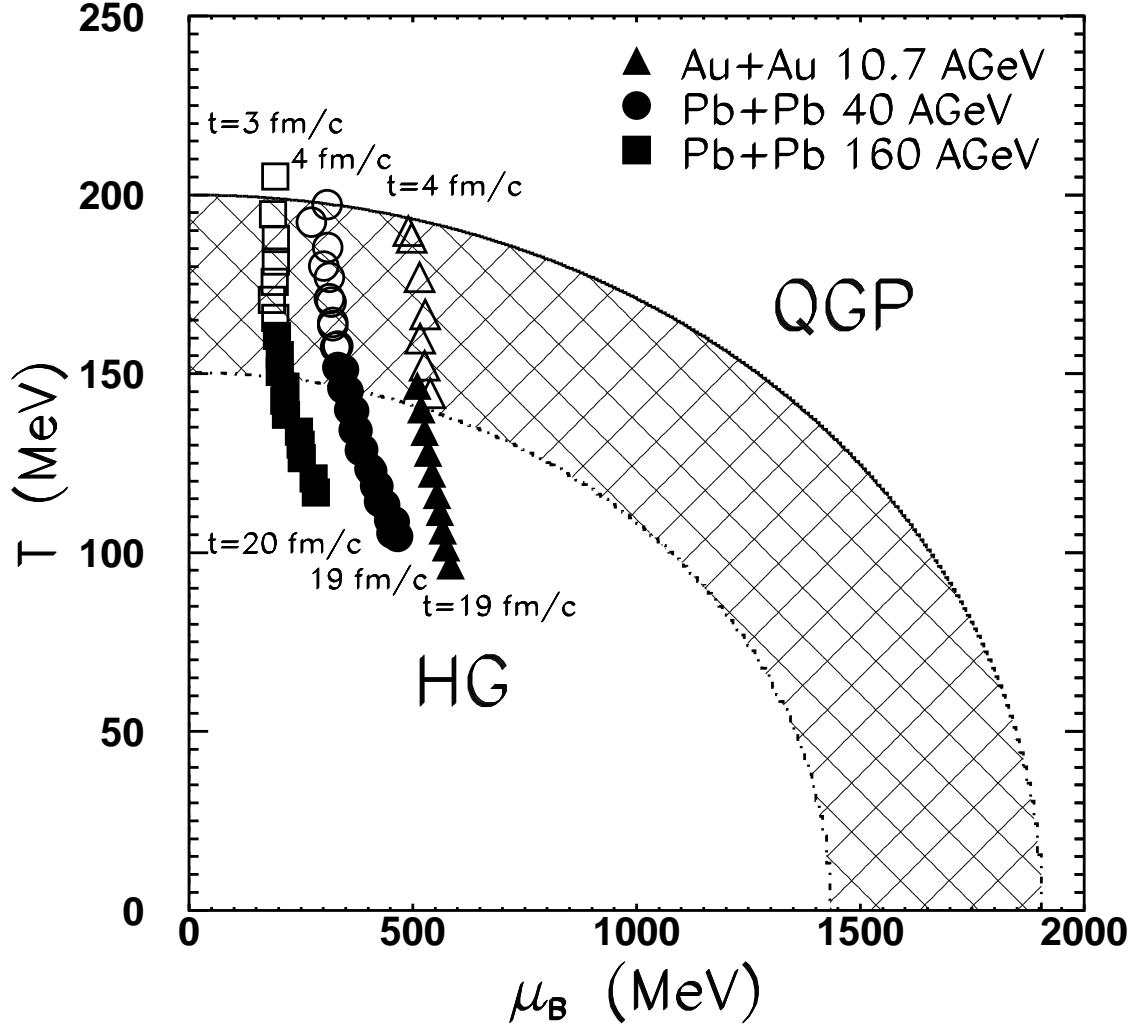}}
\caption{
The same as Fig.\protect~\ref{fig8}, but for the temperature- 
baryochemical potential ($T$-$\mu_{\rm B}$) plane. 
Both parameters are extracted from the fit to the SM.
The solid lines correspond to the boundary of the QGP calculated
for the bag constant 
$\ds B = P_{\rm QGP} = T_c^4 \left[ \frac{95}{180}\pi^2
+ \frac{1}{9} \left( \frac{\mu_{\rm B}}{T_c} \right)^2 +
\frac{1}{162} \left( \frac{\mu_{\rm B}}{T_c} \right)^4 \right]$. For
$\mu_{\rm B} = 0$ this gives us $B^{1/4} = 227$ MeV and 
302 MeV at $T_c = 150$ MeV and 200 MeV, correspondingly.
Open symbols correspond to the nonequilibrium stage of the reaction,
while full symbols indicate the stage of kinetically equilibrated 
matter.
}
\label{fig10}
\end{figure}

The evolution of the central cell in $T$-$\mu_{\rm B}$ plane is shown
in Fig.~\ref{fig10}. Here, in spite of the absence of indication on
the QGP-hadrons phase transition in $P$-$\varepsilon$ plane, we
see that the maximal temperature is growing with the initial collision
energy and reaches at the beginning of the kinetic equilibrium stage 
the zone of the phase transition predicted by the MIT bag model for
ideal QGP phase with $m_{\rm S} = 0$.
At earlier times the determination of temperature in the cell by means
of the SM fit is doubtful, since the necessary conditions of local 
equilibrium are not satisfied.   

\subsection{Hadron yields and energy spectra}
\label{subs4c}

\begin{figure}[htp]
\centerline{\epsfysize=16cm \epsfbox{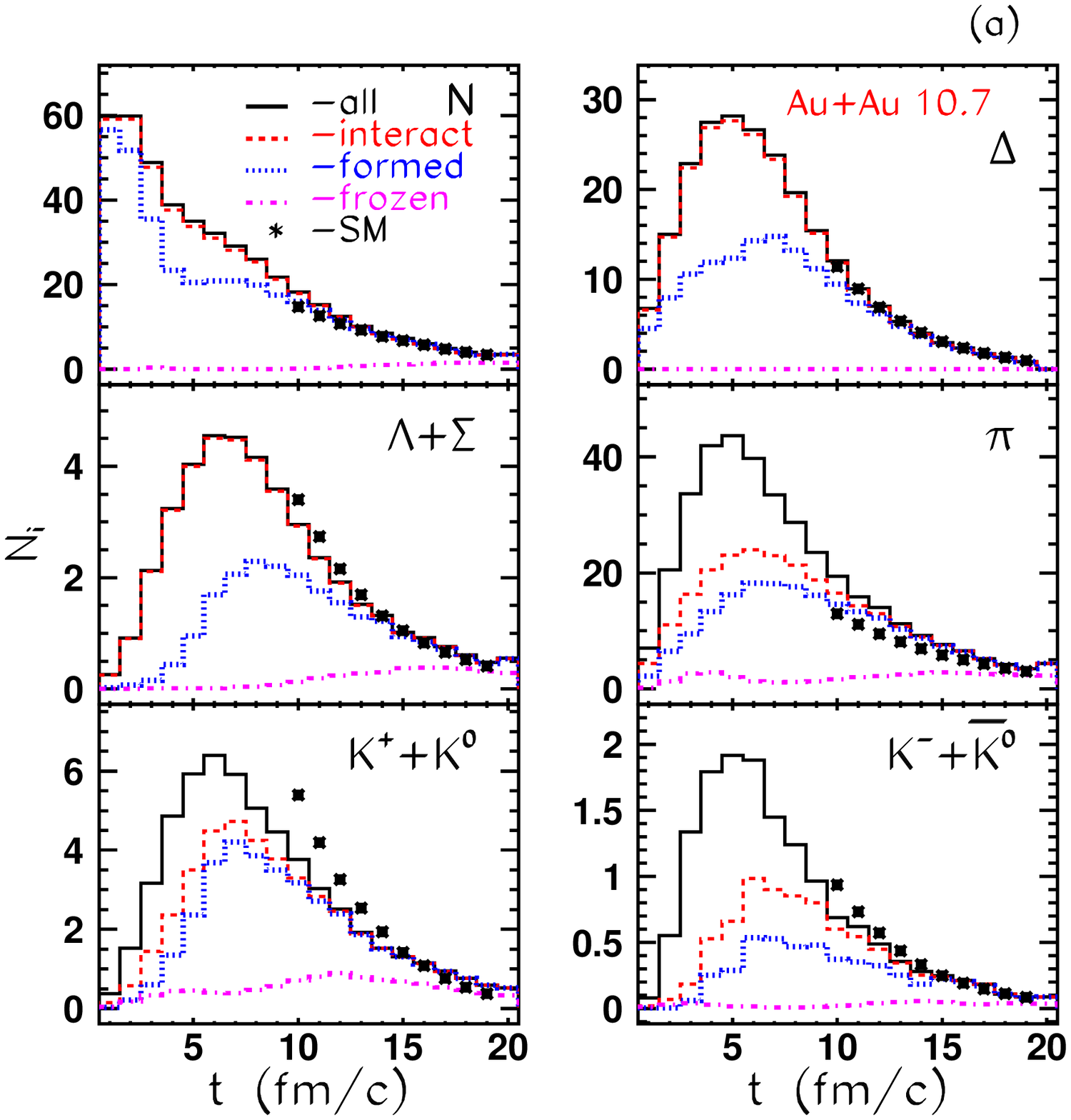}}
\caption{
(a) The number of particles in the central cell of heavy ion 
collisions at 10.7{\it A} GeV as a function of time as obtained in 
UrQMD model (histograms) together with the predictions of the SM 
(full symbols). Solid lines correspond to all hadrons in the cell,
dashed lines correspond to interacting particles, and dotted lines 
correspond to formed hadrons. The numbers of frozen particles in the 
cell are shown by dot-dashed lines.
Note that baryons can interact immediately after collision due
to their leading diquark content, while mesons can interact only
after certain formation time.\\ 
(b) The same as (a) but for Pb+Pb collisions at 40{\it A} GeV.\\
(c) The same as (a) but for Pb+Pb collisions at 160{\it A} GeV.
}
\centerline{\epsfysize=16cm \epsfbox{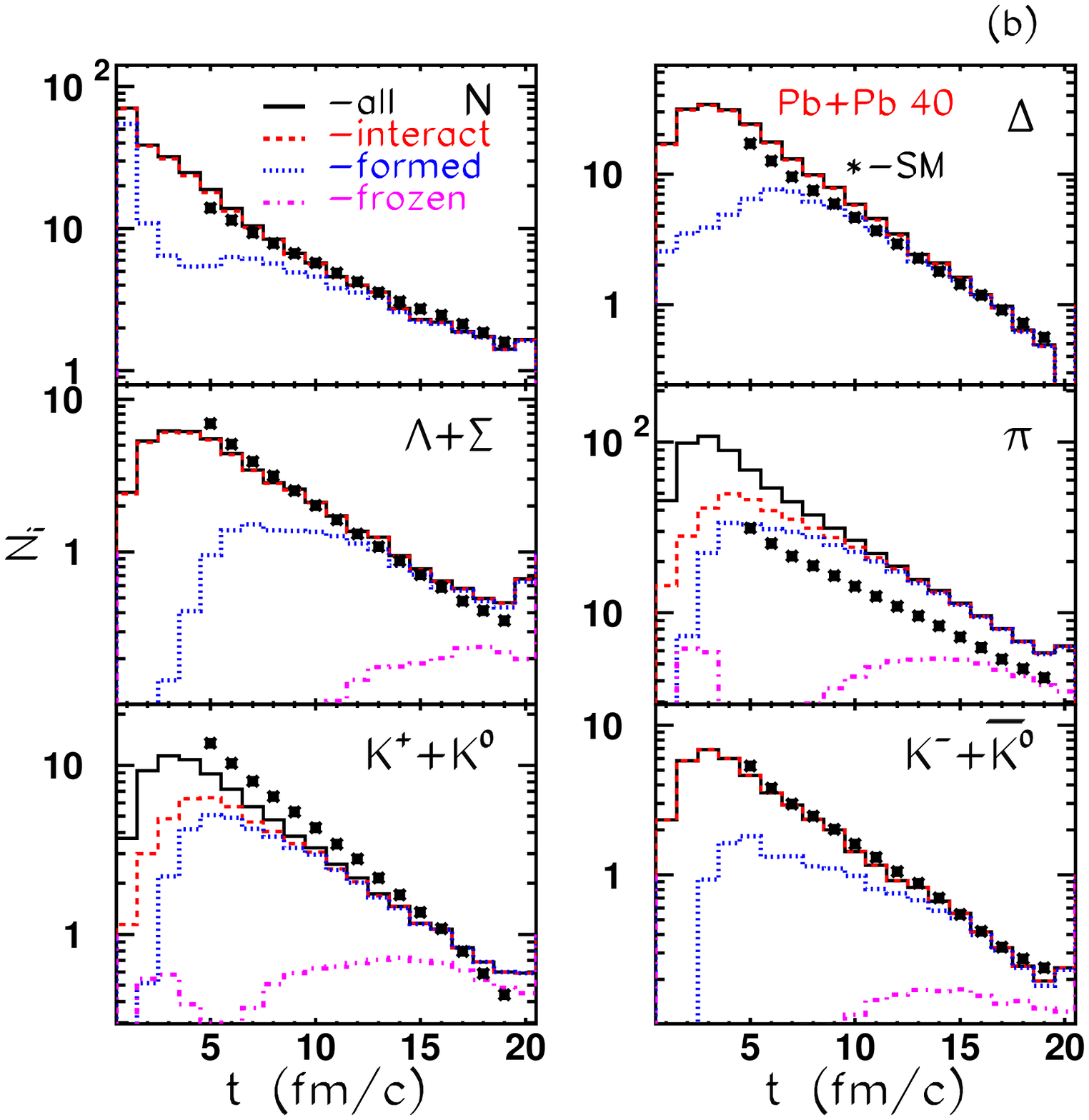}}
\centerline{\epsfysize=16cm \epsfbox{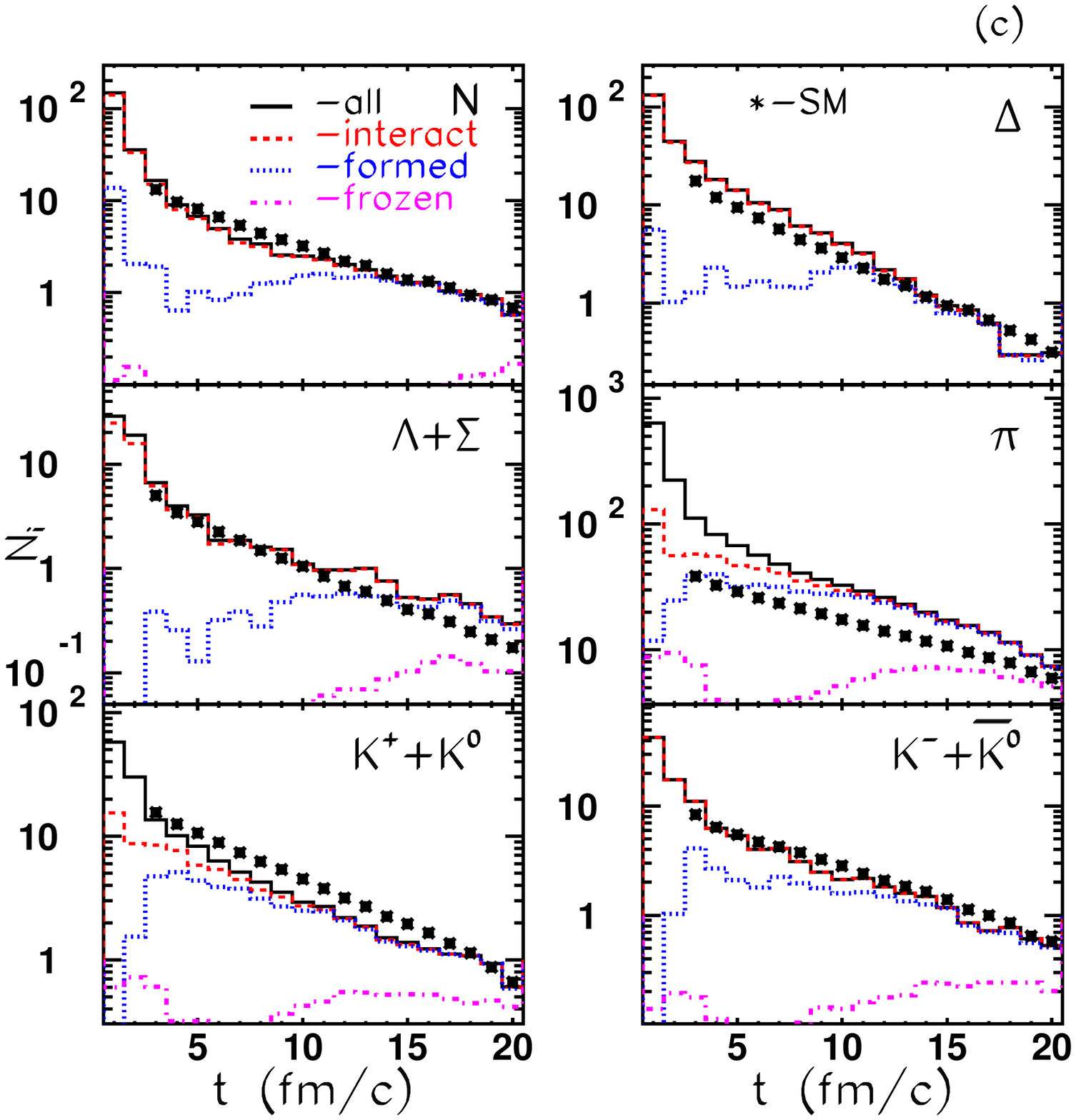}}
\label{fig11}
\end{figure}

\begin{figure}
\centerline{\epsfysize=17cm \epsfbox{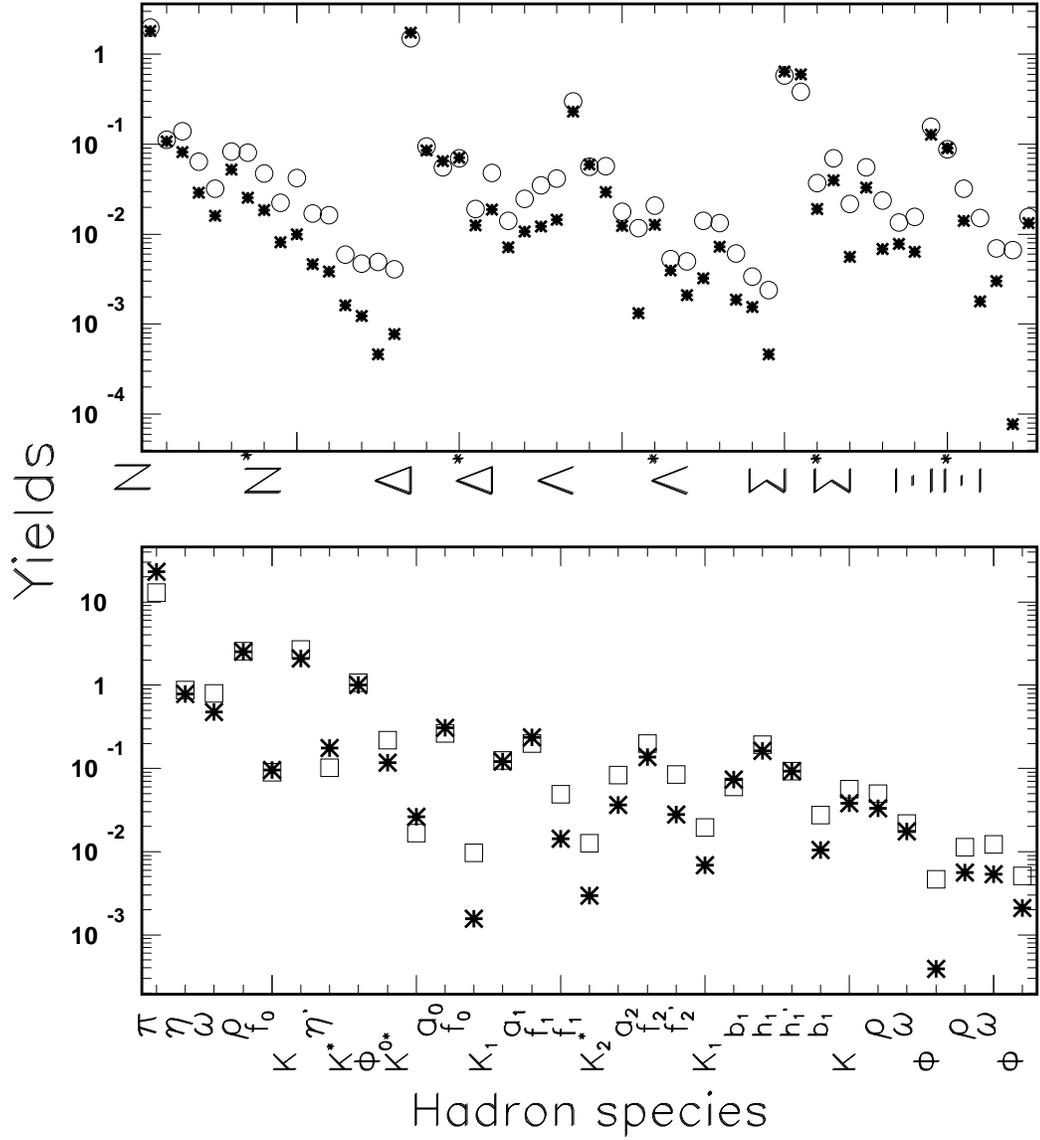}}
\caption{
Baryonic ($\bigcirc$) and mesonic ($\Box$) yields of particles 
produced at $t = 13$ fm/$c$ in Pb+Pb collisions at 160{\it A} GeV 
compared to SM predictions ($\ast$). 
}
\label{fig12}
\end{figure}

To complete the analysis of local equilibrium in the central cell
we should make a comparison between hadron yields and spectra 
obtained in the both models.
If the number of particles in the cell and their energy spectra are
very close to those predicted by the SM, one may conclude that the 
thermal and chemical equilibrium is reached. The yields of different 
hadrons in the central $V = 125$ fm$^3$ cell are shown in 
Figs.~\ref{fig11}(a)--(c) (see also Table \ref{tab4}) for central 
(impact parameter $b = 0$) heavy ion collisions at 10.7{\it A} GeV, 
40{\it A} GeV and 160{\it A} GeV, respectively. 
We see that for baryons at $t \geq 10$ fm/$c$ the agreement between
the SM and UrQMD results is reasonably good. For pions and kaons the
yields differ drastically, especially at 160{\it A} GeV. Compared to 
UrQMD, the statistical model significantly underestimates the number
of pions and overestimates the kaon yield. This difference in the pion
yield cannot be explained only by the many-body ($N \geq 3$) decays 
of resonances, whose number is lower in the UrQMD calculations as seen 
in Fig.~\ref{fig12}. In fact, the main discrepancy is observed for 
pions and many-body decaying resonances, like $\omega, N^{\ast}, 
\Delta^{\ast}, \Lambda^{\ast}$, etc. Statistical model overestimates 
the production of such resonances and underestimates the yield of 
pions. The enhancement of the resonances, however, can describe only 
20\% of difference in pion yields, the other 80\% are coming via the 
multiparticle processes, i.e., fragmentations of strings.
The condition {\bf (iv)} is not satisfied and, therefore, the hadronic
matter in the UrQMD central cell is not chemically equilibrated.

\begin{figure}
\centerline{\epsfysize=17cm \epsfbox{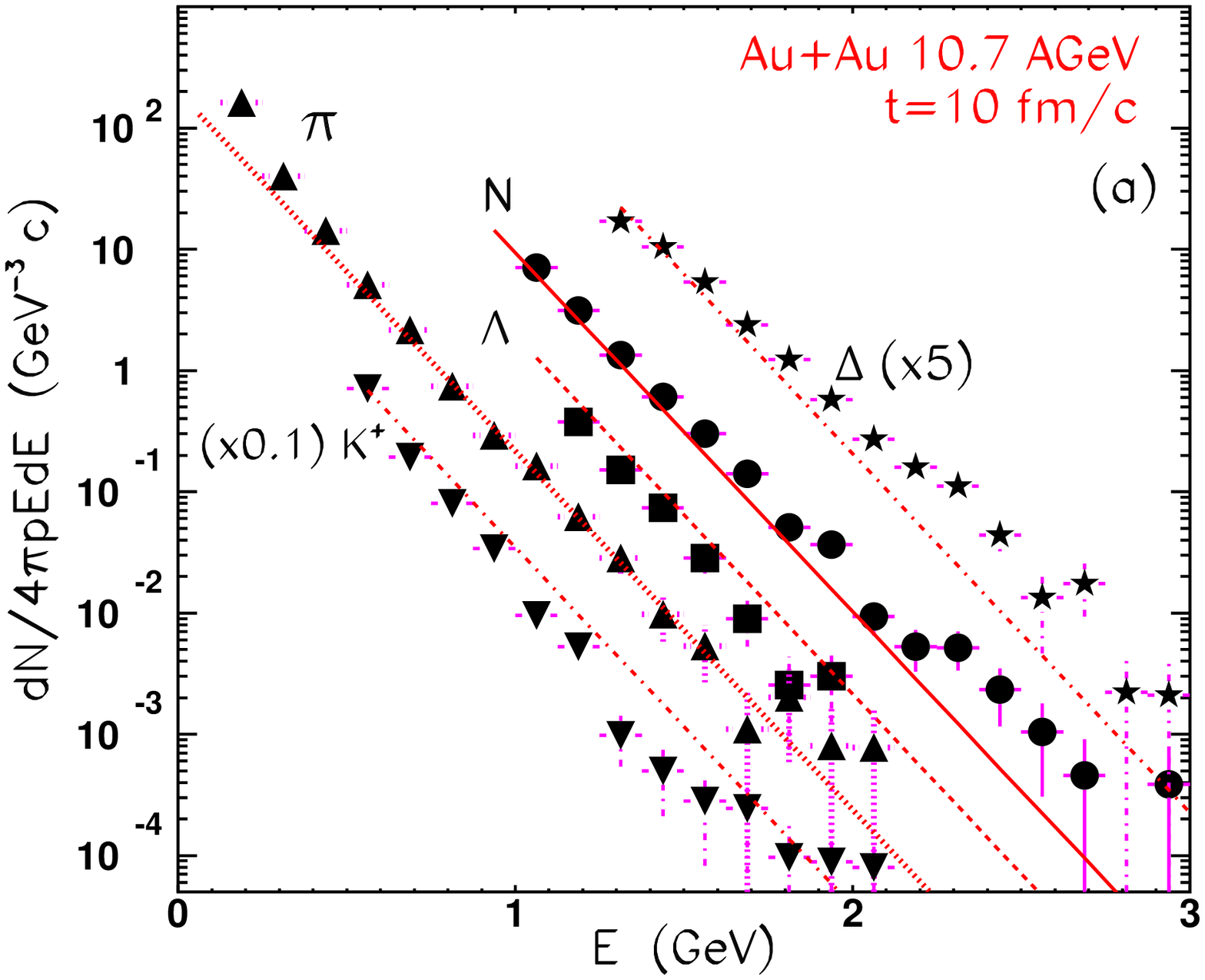}}
\caption{
(a) Energy spectra of $N$ ($\bullet $), $\Lambda$ ($\Box $),
$\pi$ ($\bigtriangleup $), $K^+$ ($\bigtriangledown $), 
$K^-$ ($\ast$), and $\Delta$ ($\star$) in the central 125 fm$^3$ cell 
of Au+Au collisions at 10.7{\it A} GeV at $t$=10~fm/$c$.
Dashed lines are the results of Boltzmann fit to the distributions
with the parameters $T$=147 MeV, $\mu_B$=510 MeV, and $\mu_S$=129 MeV
obtained in the ideal hadron gas model.\\
(b) The same as (a) but for Pb+Pb collisions at 40{\it A} GeV.
Parameters of the Boltzmann fit are $T$=151 MeV,
$\mu_B$=345 MeV, and $\mu_S$=74 MeV.\\
(c) The same as (a) but for Pb+Pb collisions at 160{\it A} GeV.
Parameters of the Boltzmann fit are $T$=161 MeV,
$\mu_B$=197 MeV, and $\mu_S$=36.8 MeV.\\
}
\centerline{\epsfysize=17cm \epsfbox{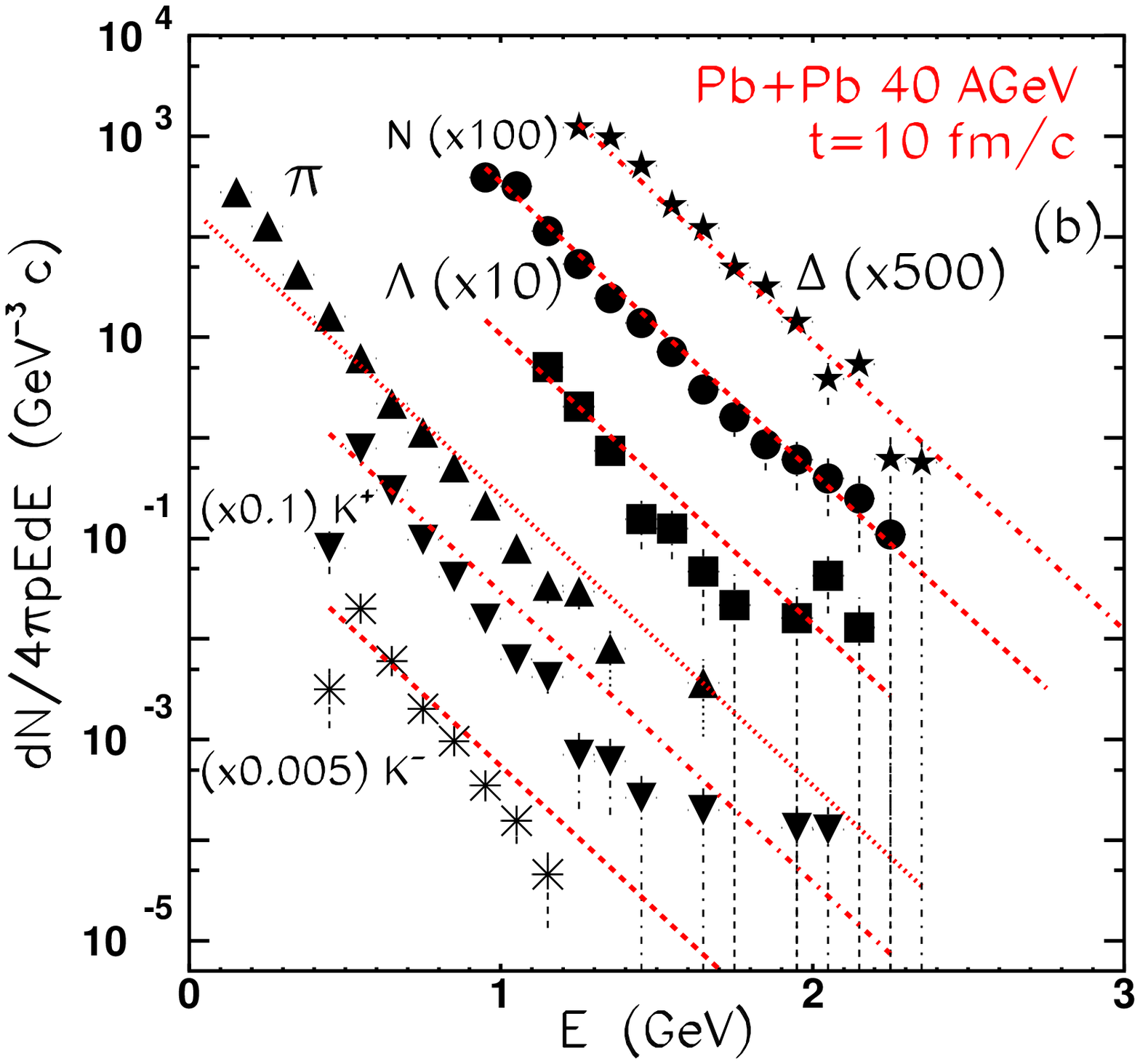}}
\centerline{\epsfysize=17cm \epsfbox{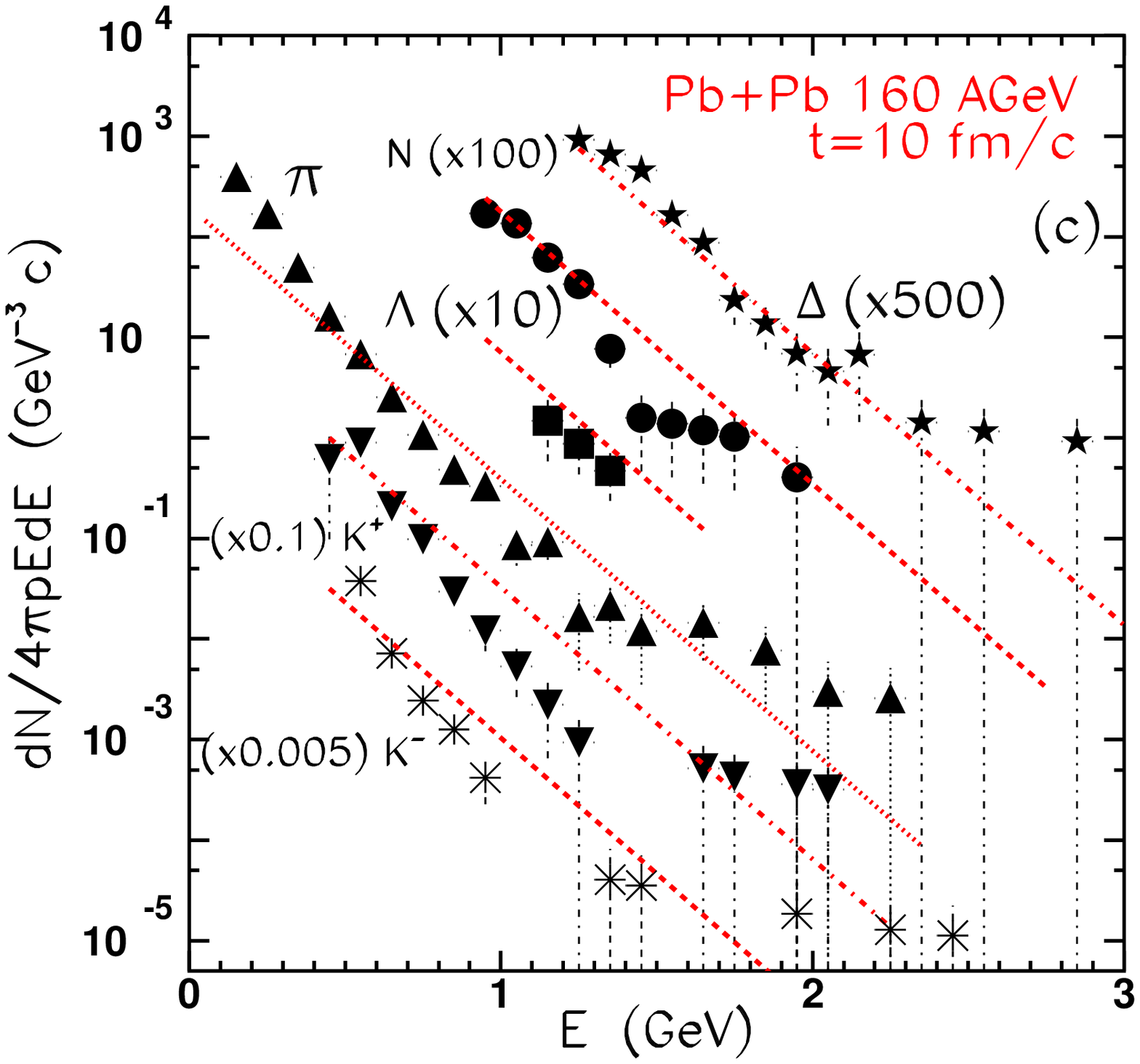}}
\label{fig13}
\end{figure}

To verify how good the SM reproduces the temperature of the system
we display in Figs.~\ref{fig13}(a)-\ref{fig13}(c) the energy spectra 
of different
hadronic species, obtained from the microscopic calculations. The
predictions of the statistical model are plotted onto the particle
spectra, too. Again, at AGS energy the difference between the UrQMD
and SM results for baryons lies within the 10\%- range of accuracy.
With the rise of initial energy from AGS to SPS the agreement between
the models in the baryonic sector becomes worse. Pion energy spectra
demonstrate the same tendency. Moreover, even at 10.7{\it A} GeV the 
deviations of pion spectra in UrQMD from those of the SM are 
significant. The Boltzmann fit to pion and nucleon energy spectra 
from the central cell has been performed at 160{\it A} GeV, where the 
deviations from the SM predictions are especially noticeable. Results
of the fit are listed in Table~\ref{tab5}. We see that the nucleon
``temperature" is always 30--40 MeV below the temperature obtained 
in the statistical model. For pions the difference is more dramatic:
50--60 MeV, although the UrQMD energy spectra themselves agree well
with the exponential form of Boltzmann distribution. 

\begin{figure}
\centerline{\epsfysize=17cm \epsfbox{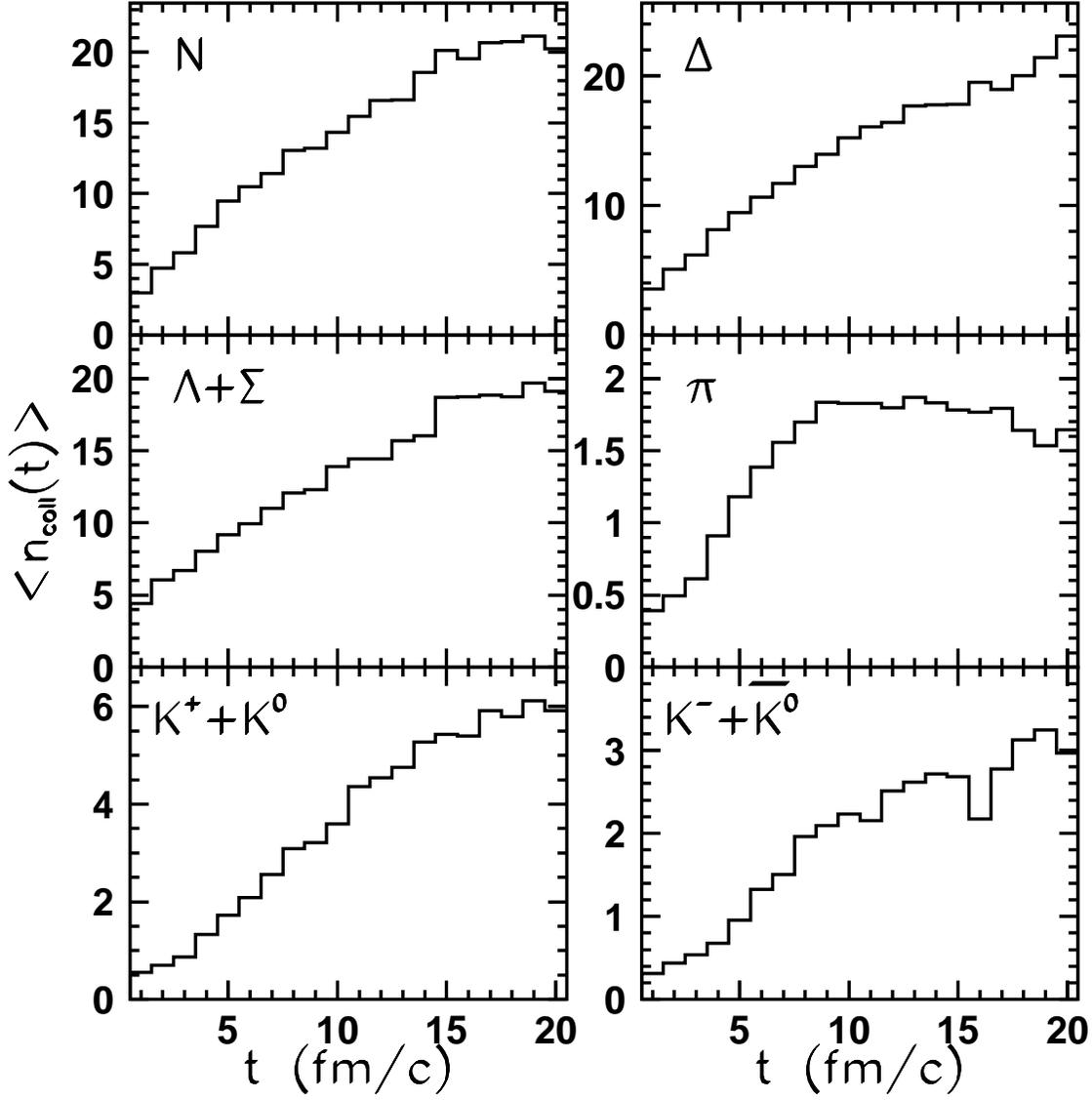}}
\caption{
Time evolution of average number of collisions 
for $N$, $\Delta$, $\Lambda + \Sigma$, $\pi$, $K^+ + K^0$, and
$K^- + \overline{K^0}$ in the central cell of Pb+Pb collisions 
at 160{\it A} GeV.
}
\label{fig14}
\end{figure}

But maybe all these nonequilibrium effects are caused solely by
pions which, due to their simultaneous production in inelastic
collisions and decays of resonances, are the only hadrons not
thermally and chemically equilibrated? Indeed, from Fig.~\ref{fig14},
which depicts the evolution of the average number of collisions per
particle in the cell at SPS energy, it follows that pions have 
undergone about 1.6--1.7 elastic collisions while baryons have 
suffered more than 20 strong interactions. 
Thus, it would be expected that without pions the SM will predict 
much lower temperature which will agree
with that of the UrQMD. To check this hypothesis we subtract the 
energy in the cell carried by pions from the total energy of hadrons.
Then we substitute the new value of the energy density together with
the unchanged values of baryon and strangeness densities into the SM
fit to the UrQMD data, and impose the requirement of absence of pions. 
Results of the fit are listed in Table~\ref{tab6} for Pb+Pb at 
160{\it A} GeV at $t = 10$ fm/$c$. Although the number of pions in 
the cell is almost two times larger than that of SM, it appears that,
due to lower temperature of pions in UrQMD, the total excess of pion
energy density in the cell is 24 MeV/fm$^3$, or about $1/3$ of the 
total pion energy density given by the statistical model. The
contamination of pion fraction does not decrease the temperature in 
the SM. Instead, it leads to the increase of chemical potential of 
strange particles.

Therefore, despite the occurrence of a state in which hadrons are
in kinetic equilibrium and collective flows are very small, the 
hadronic matter is neither in thermal nor in chemical equilibrium.
This state of hot hadronic matter is very peculiar \cite{irr99} and 
the results of the investigations will be published elsewhere 
\cite{part2}. Similar results have been obtained in \cite{GeKa93},
where the central region of ultra-relativistic Au+Au collisions at
RHIC energy was studied using the parton cascade model \cite{Geig95}.
It was found that, despite approaching kinetic equilibrium in the
system, the chemical composition of quarks and gluons was not in
chemical equilibrium. Our analysis shows also that the
extraction of temperature by performing the SM fit to hadron yields 
and energy spectra is a very delicate procedure. If the whole system 
is out of the equilibrium state, than the ``apparent" temperatures 
obtained from the fit may occur high enough to hit the zone of 
quark-hadron phase transition or even pure QGP phase 
(Fig.~\ref{fig10}, open symbols).
 
\section{Conclusions}
\label{sec5}

The results of the present study may be summarized as follows.
We used the microscopic transport UrQMD model to verify the 
appearance of the local equilibrium in the central zone of heavy ion
collisions at relativistic energies, spanning from AGS to SPS. To 
analyze the results of the
dynamical calculations the traditional methodic has been applied.
First, the conditions of preequilibrium kinetic stage have been
checked by means of the isotropy of the pressure and velocity
distributions. It is shown that the kinetic equilibrium is reached
by hadronic matter in the central $V = 125$ fm$^3$ cell at about
$t = 10$ fm/$c$ for a not very long period, $\Delta t \cong 8 - 10$
fm/$c$. Secondly, the values of the energy density, baryonic
density and strangeness density, calculated microscopically, were
used as an input to calculate temperature as well as baryonic and 
strangeness chemical potentials within the statistical model of
ideal hadron gas.

The total strangeness of all hadronic species carrying strangeness 
charge in the central cell is shown to be negative though small. This 
is because of the fact that $K$'s escape from the interaction zone
much easier than $\Lambda$'s or $\overline{K}$'s due to their small 
interaction cross section with hadrons. The small negative strangeness 
of the central cell, however, cannot be neglected because it affects 
the ratios of strange particles, like $K/\overline K$.
 
It is worth to note that due to rather complicated dynamics of heavy
ion collisions thermal models cannot fully describe the bulk of 
experimental data \cite{Hein98}. In contrast, in the symmetric central 
zone of the heavy ion collisions almost all dynamical factors are 
reduced. This gives us a chance to study the relaxation of the hot
hadronic matter to the thermal and chemical equilibrium, provided
it would set in within the hadronization time of the system.

We found that the entropy per baryon in the central cell remains 
constant at the late stage of the expansion for energies varying from 
10.7{\it A} GeV to 160{\it A} GeV. This circumstance formally 
supports the application of the relativistic hydrodynamical model.
But the further comparison between the predictions of the microscopic  
and macroscopic models reveals significant discrepancies in the
yields and energy spectra of hadrons. Compared to UrQMD, the 
statistical model underestimates, for instance, the number of pions.
This ``meson problem" is not a feature attributed solely to the 
particular microscopic model like UrQMD. Experimental data on pion
yields at SPS energies \cite{NA49} show unambiguously the enhancement 
of pions compared to the SM calculations. Several possible solutions 
have been suggested recently. Admitting that the hot hadronic matter 
appears not in the state of chemical equilibrium, one may implement 
the effective chemical potential for pions (and other species, too) 
\cite{YeGo98,LeRa98}. In that case the state of maximum entropy is 
not reached \cite{LeRa98} yet.

Also, the temperatures of different hadronic species are not the same. 
The differences between the UrQMD and SM results increase with the 
rise of initial energy of colliding nuclei. Pions seem to have the 
lowest temperature and nucleons the highest one among all hadron 
species. Both, the pion and nucleon temperatures in the cell in Pb+Pb 
collisions at SPS energy are always far below the
temperature predicted by the statistical thermal model.

The information at hand permits us to summarize that, in contrast
to low energies, local thermal and chemical equilibrium (in the sense 
of the thermal model) is not obtained even in the central zone of 
heavy ion collisions at energies above 10.7{\it A} GeV in the UrQMD
simulations. The hadronic matter in the UrQMD model seems not to 
evolve towards the state of maximum entropy, and this fact deserves 
to be investigated in detail.    

\section*{Acknowledgments} 
We would like to thank L. Csernai, U. Heinz, J. Rafelski, L. Satarov,
E. Shuryak, J. Stachel, and N. Xu for the helpful discussions and
comments.
L.B. and E.Z. are grateful to the Institute for Theoretical Physics, 
University of Frankfurt for the warm and kind hospitality. 
This work was supported by the Graduiertenkolleg f{\"u}r Theoretische
und Experimentelle Schwerionenphysik, Frankfurt--Giessen, the
Bundesministerium f{\"u}r Bildung und Forschung, the Gesellschaft
f{\"u}r Schwerionenforschung, Darmstadt, Deutsche
Forschungsgemeinschaft, and the Alexander von Humboldt--Stiftung,
Bonn.

\newpage

\newpage
\mediumtext

\begin{table}
\caption{
The time evolution of the thermodynamic characteristics of hadronic 
matter in the central cell of volume $V = 125$ fm$^3$ in central Au+Au 
collisions at AGS (10.7{\it A} GeV) energy. The temperature, $T$, 
baryochemical potential, $\mu_{\rm B}$, strange chemical potential,
$\mu_{\rm S}$, pressure, $P$, entropy density, $s$, and entropy 
density per baryon, $s/\rho_{\rm B}$, are extracted from the 
statistical model of ideal hadron gas, using the microscopically
evaluated energy density, $\varepsilon^{\rm cell}$, baryonic density,
$\rho_{\rm B}^{\rm cell}$, and strangeness density, 
$\rho_{\rm S}^{\rm cell}$, as input. 
}

\begin{tabular}{cccccccccc}
Time & $\varepsilon^{\rm cell}$ & $\rho_{\rm B}^{\rm cell}$ &
$\rho_{\rm S}^{\rm cell}$ & $T$ & $\mu_{\rm B}$ & $\mu_{\rm S}$ & 
$P$ & $s$ & $s/\rho_{\rm B}^{\rm cell}$ \\
fm/$c$ & MeV/fm$^3$ & fm$^{-3}$ & fm$^{-3}$ & MeV & MeV & MeV &
 MeV/fm$^3$ & fm$^{-3}$ &  \\
\tableline\tableline
10 & 674.8  & 0.332  & -0.0078 & 128.46 & 146.74 & 510.03 &
  82.67 & 4.01 & 12.07 \\
11 & 510.7  & 0.261  & -0.0070 & 122.06 & 140.15 & 519.16 &
  61.99 & 3.12 & 11.95 \\
12 & 384.4  & 0.204  & -0.0055 & 116.37 & 133.87 & 526.73 &
  46.41 & 2.42 & 11.90 \\
13 & 293.4  & 0.160  & -0.0037 & 112.38 & 127.94 & 534.51 &
  35.06 & 1.90 & 11.86 \\
14 & 222.5  & 0.125  & -0.0028 & 107.37 & 122.12 & 542.13 &
  26.35 & 1.48 & 11.82 \\
15 & 170.5  & 0.100  & -0.0029 & 100.73 & 116.22 & 553.56 &
  19.88 & 1.16 & 11.62 \\
16 & 134.0  & 0.081  & -0.0024 & 95.94  & 111.42 & 560.33 &
  15.47 & 0.94 & 11.59 \\
17 & 102.5  & 0.063  & -0.0026 & 86.85  & 106.28 & 566.99 &
  11.69 & 0.74 & 11.56 \\
18 & 79.56  & 0.051  & -0.0024 & 78.90  & 101.46 & 574.17 &
   8.83 & 0.58 & 11.55 \\
\end{tabular}
\label{tab1}
\end{table}

\begin{table}
\caption{
The same as Table \protect\ref{tab1} but for central Pb+Pb collisions
at 40{\it A} GeV.
}

\begin{tabular}{cccccccccc}
Time & $\varepsilon^{\rm cell}$ & $\rho_{\rm B}^{\rm cell}$ &
$\rho_{\rm S}^{\rm cell}$ & $T$ & $\mu_{\rm B}$ & $\mu_{\rm S}$ &
$P$ & $s$ & $s/\rho_{\rm B}^{\rm cell}$ \\
fm/$c$ & MeV/fm$^3$ & fm$^{-3}$ & fm$^{-3}$ & MeV & MeV & MeV &
 MeV/fm$^3$ & fm$^{-3}$ &  \\
\tableline\tableline
10 & 428.6 & 0.146 & -0.0063 & 151.11 & 344.97 & 74.03 & 60.47 &
  2.90 & 19.76 \\
11 & 327.9 & 0.116 & -0.0056 & 145.15 & 355.60 & 69.72 & 46.44 &
  2.29 & 19.78 \\
12 & 253.3 & 0.093 & -0.0038 & 139.65 & 367.50 & 68.41 & 36.00 & 
  1.83 & 19.77 \\
13 & 195.3 & 0.073 & -0.0047 & 134.13 & 375.42 & 60.13 & 27.93 &
  1.46 & 19.97 \\
14 & 151.6 & 0.059 & -0.0036 & 128.77 & 388.19 & 57.76 & 21.74 & 
  1.17 & 19.95 \\
15 & 118.9 & 0.048 & -0.0030 & 123.45 & 404.47 & 55.82 & 17.01 & 
  0.94 & 19.65 \\
16 &  95.4 & 0.040 & -0.0022 & 118.52 & 422.08 & 56.00 & 13.55 & 
  0.78 & 19.21 \\
17 &  74.0 & 0.033 & -0.0024 & 113.26 & 437.23 & 49.81 & 10.46 & 
  0.62 & 18.94 \\
18 &  59.88& 0.027 & -0.0028 & 109.06 & 447.99 & 41.27 &  8.43 & 
  0.51 & 18.82 \\ 
\end{tabular}
\label{tab2}
\end{table}

\begin{table}
\caption{
The same as Table \protect\ref{tab1} but for central Pb+Pb collisions
at SPS (160{\it A} GeV) energy.
}

\begin{tabular}{cccccccccc}
Time & $\varepsilon^{\rm cell}$ & $\rho_{\rm B}^{\rm cell}$ &
$\rho_{\rm S}^{\rm cell}$ & $T$ & $\mu_{\rm B}$ & $\mu_{\rm S}$ &
$P$ & $s$ & $s/\rho_{\rm B}^{\rm cell}$ \\
fm/$c$ & MeV/fm$^3$ & fm$^{-3}$ & fm$^{-3}$ & MeV & MeV & MeV &
 MeV/fm$^3$ & fm$^{-3}$ &  \\
\tableline\tableline
10 & 467.0  & 0.092 & -0.0099 & 160.564 & 196.64 & 36.78 & 71.91 &
  3.24 & 35.13 \\
11 & 364.0  & 0.073 & -0.0077 & 155.208 & 202.76 & 34.73 & 56.76 &
  2.61 & 35.83 \\
12 & 287.0  & 0.056 & -0.0057 & 150.467 & 202.90 & 32.04 & 45.55 &
  2.13 & 37.85 \\
13 & 239.0  & 0.049 & -0.0067 & 146.453 & 215.20 & 28.71 & 38.29 &
  1.82 & 36.86 \\
14 & 193.0  & 0.039 & -0.0067 & 142.419 & 212.18 & 22.27 & 31.46 &
  1.51 & 39.41 \\
15 & 158.0  & 0.031 & -0.0044 & 138.531 & 216.93 & 23.23 & 26.19 &
  1.28 & 41.04 \\
16 & 130.0  & 0.029 & -0.0037 & 133.837 & 245.73 & 25.20 & 21.51 &
  1.08 & 37.56 \\
17 & 106.9  & 0.023 & -0.0037 & 130.165 & 249.13 & 20.46 & 17.96 &
  0.91 & 39.14 \\
18 &  86.9  & 0.018 & -0.0029 & 126.450 & 251.17 & 18.34 & 14.89 &  
  0.77 & 41.68 \\
\end{tabular}
\label{tab3}
\end{table}

\begin{table}
\caption{
Hadron yields (without resonance feeding) in the central cell 
calculated for all three reactions at $t = 10$ fm/$c$ within UrQMD 
and SM with nonzero and zero strangeness density. Value of the 
nonzero strangeness density is taken from the UrQMD simulations.
}

\begin{tabular}{cccccccccc}
Hadrons & \multicolumn{3}{c}{ 10.7{\it A} GeV} & 
 \multicolumn{3}{c}{ 40{\it A} GeV} & 
 \multicolumn{3}{c}{ 160{\it A} GeV} \\ 
\cline{2-10}
 & \multicolumn{2}{c} {SM} & UrQMD & \multicolumn{2}{c} {SM} & UrQMD
 & \multicolumn{2}{c} {SM} & UrQMD \\  
 & $\ \rho_{\rm S}$=0 & $\ \rho_{\rm S} = 
         \rho_{\rm S}^{\rm cell}$ & $ \rho_{\rm S}^{\rm cell}$ &
   $\ \rho_{\rm S}$=0 & $\ \rho_{\rm S} = 
         \rho_{\rm S}^{\rm cell}$ & $ \rho_{\rm S}^{\rm cell}$ &
   $\ \rho_{\rm S}$=0 & $\ \rho_{\rm S} = 
         \rho_{\rm S}^{\rm cell}$ & $ \rho_{\rm S}^{\rm cell}$ \\
\tableline \tableline
$\pi$                &12.90 &12.99 & 19.40 
     &14.35 &14.41 & 26.65 & 17.43 &17.47 & 33.02   \\
$N$                  &14.93 &14.70 & 18.19
     & 5.81 & 5.69 &  5.72 &  3.28 & 3.20 &  2.53  \\
$\Delta $            &11.53 &11.40 & 12.08
     & 4.73 & 4.64 &  5.86 &  2.97 & 2.89 &  4.17  \\
$\Lambda+\Sigma $    & 3.24 & 3.40 & 2.95
     & 1.94 & 2.03 &  2.11 &  1.51 & 1.58 &  1.17  \\
$\Lambda^*+\Sigma^*$ & 3.06 & 3.22 & 2.72
     & 1.98 & 2.08 &  1.94 &  1.79 & 1.88 &  1.90  \\
$\Xi +\Xi^*$         & 0.54 & 0.60 & 0.21
     & 0.51 & 0.57 &  0.31 &  0.57 & 0.64 &  0.37  \\
$K^+ + K^0$          & 5.68 & 5.40 & 3.77
     & 4.54 & 4.28 &  3.24 &  4.84 & 4.50 &  2.98  \\
$K^{+*} + K^{0*}$    & 2.98 & 2.85 & 2.31
     & 2.65 & 2.50 &  2.65 &  3.39 & 3.18 &  3.40  \\
$K^- +\overline K^0$ & 0.88 & 0.94 & 0.69
     & 1.53 & 1.65 &  1.43 &  2.64 & 2.84 &  2.17  \\
$K^{-*}+\overline K^{0*}$ & 0.48 & 0.52 & 0.30
     & 0.91 & 0.98 &  0.76 &  1.88 & 2.03 &  1.95  \\
$\overline\Lambda+\overline\Sigma$    &0.19 & 0.19 & 0.00
     &0.059 &0.058 & 0.013 &0.223 &0.216 & 0.13 \\
$\overline\Lambda^*+\overline\Sigma^*$&0.18 & 0.18 & 0.00
     &0.060 &0.059 & 0.025 &0.264 &0.256 & 0.10 \\
\end{tabular}
\label{tab4}
\end{table}

\begin{table}
\caption{
The temperature, $T_{\rm SM}^{\rm all}$, extracted from the SM fit to 
UrQMD data at given $\varepsilon,\ \rho_{\rm B}$, and $\rho_{\rm S}$,
together with the temperature of nucleons, $T^N_{\rm fit}$, and
pions, $ T^{\pi}_{\rm fit}$, obtained by the Boltzmann fit to energy
spectra of particles at 160{\it A} GeV within the time interval 
$10 \leq t \leq 15$ fm/$c$.
}

\begin{tabular}{cccc}
Time & $T^{\rm all}_{SM}$ & $ T^N_{\rm fit}$ & $ T^{\pi}_{\rm fit}$ \\
fm/$c$ &  MeV  &  MeV  &  MeV  \\
\tableline\tableline
10 & 160.56  &  $ 121 \pm 0.1 $  &  $ 99 \pm 0.1 $ \\
11 & 155.21  &  $ 114 \pm 0.1 $  &  $ 96 \pm 0.1 $ \\
12 & 150.47  &  $ 108 \pm 0.1 $  &  $ 92 \pm 0.1 $ \\
13 & 146.45  &  $ 121 \pm 0.1 $  &  $ 89 \pm 0.1 $ \\
14 & 142.42  &  $ 113 \pm 0.1 $  &  $ 86 \pm 0.1 $ \\
15 & 138.53  &  $ 108 \pm 0.1 $  &  $ 85 \pm 0.1 $ \\
\end{tabular}
\label{tab5}
\end{table}

\begin{table}
\caption{
The temperature, $T$, baryochemical potential, $\mu_{\rm B}$, and
strange chemical potential, $\mu_{\rm S}$, extracted from the SM
fit to the energy density, $\varepsilon^{\rm cell}$, baryonic density,
$\rho_{\rm B}^{\rm cell}$, and strangeness density, 
$\rho_{\rm S}^{\rm cell}$, with (upper line) and without (bottom line)
pionic fraction in Pb+Pb collisions at 160{\it A} GeV at time 
$t = 10$ fm/$c$.
}

\begin{tabular}{ccccccc}
 & $\varepsilon^{\rm cell}$ & $\rho_{\rm B}^{\rm cell}$ &
$\rho_{\rm S}^{\rm cell}$ & $T$ & $\mu_{\rm B}$ & $\mu_{\rm S}$ \\
 & MeV/fm$^3$ & fm$^{-3}$ & fm$^{-3}$ & MeV & MeV & MeV  \\
\tableline\tableline
{\rm With pions}    & 467.0 & 0.0924 & -0.00986 & 160.56 & 196.64 & 
36.78 \\
{\rm Without pions} & 369.0 & 0.0924 & -0.00986 & 160.50 & 195.50 & 
48.61 \\
\end{tabular}
\label{tab6}
\end{table}

\end{document}